\documentclass{aa} 
\usepackage{natbib}
\usepackage{amsmath}
\usepackage{psfig}
\usepackage{longtable,lscape}
\bibpunct{(}{)}{;}{a}{}{,} %Astronomy & Astrophysics, ApJ
\makeatletter
\edef\@tempa#1#2{\def#1{\mathaccent\string"\noexpand\accentclass@#2 }}
\@tempa\cdring{017}
\makeatother
\newcommand{\um}{\,\ensuremath{\mu\textrm{m}}}

\newcommand{\Mdisk}{$M_{\mathrm{d}}$}
\newcommand{\Teff}{$T_{\mathrm{eff}}$}

\newcommand{\refaccent}[1]{\cdring{#1}}

\newcommand{\Lstar}{\ensuremath{L_{\star}}}

\newcommand{\Tstar}{\ensuremath{T_{\star}}}
\newcommand{\Tdust}{\ensuremath{T_{\mathrm{d}}}}
\newcommand{\Tstarref}{\ensuremath{\refaccent{T}_{\star}}}
\newcommand{\Tdustref}{\ensuremath{\refaccent{T}_{\mathrm{d}}}}

\begin{document}
\thesaurus{N08,08.16.2,13.09.6}
\title{Incidence and survival of remnant disks around
main-sequence stars}
\authorrunning{H.J. Habing et al.}
\titlerunning{Disks surrounding main-sequence stars}
\author{H.J.~Habing\inst{1}
 \and C.~Dominik\inst{1}
 \and M.~Jourdain~de~Muizon\inst{2,3}
 \and R.J.~Laureijs\inst{4}
 \and M.F.~Kessler\inst{4}
 \and K.~Leech\inst{4}
 \and L.~Metcalfe\inst{4}
 \and A.~Salama\inst{4}
 \and R.~Siebenmorgen\inst{4}
 \and N.~Trams\inst{4}
 \and P.~Bouchet\inst{5}
}
\institute{Sterrewacht, Leiden, P.O. Box 9513, NL-2300 RA Leiden, The
  Netherlands \and DESPA, Observatoire de Paris, F-92190 Meudon, France
  \and LAEFF-INTA, ESA Vilspa, PO Box 50727, 28080 E-Madrid, Spain \and
  ISO Data Center, Astrophysics Division of ESA, Vilspa, PO Box 50727,
  E-28080 Madrid, Spain \and Cerro Tololo Inter-American Observatory,
  NOAO, Casilla 603, La Serena, Chile 1353 }
\offprints{habing@strw.leidenuniv.nl (H. J. Habing)}
\thanks{Based on observations with ISO, an ESA project with
  instruments funded by ESA Member States (especially the PI
  countries: France, Germany, The Netherlands and the United Kingdom)
  and with the participation of ISAS and NASA.}
\date{Received:  ; accepted:  }
\maketitle 
\begin{abstract}
\begin{sloppypar} We present photometric ISO 60 and 170 \um\ measurements, 
  complemented by some IRAS data at 60
  \um , of a sample of 84 nearby main-sequence stars of spectral class
  A, F, G and K in order to determine the incidence of dust disks
  around such main-sequence stars.  Fifty stars were detected at
  60\um; 36 of these emit a flux expected from their photosphere while
  14 emit significantly more. The excess emission we attribute to a
  circumstellar disk like the ones around Vega and $\beta$ Pictoris.
  Thirty four stars were not detected at all; the expected
  photospheric flux, however, is so close to the detection limit that
  the stars cannot have an excess stronger than the photospheric flux
  density at 60\,\um.
  
  Of the stars younger than 400 Myr one in two has a disk; for the
  older stars this is true for only one in ten. We conclude that most
  stars arrive on the main sequence surrounded by a disk; this disk
  then decays in about 400 Myr.  Because (i) the dust particles
  disappear and must be replenished on a much shorter time scale and
  (ii) the collision of planetesimals is a good source of new dust, we
  suggest that the rapid decay of the disks is caused by the
  destruction and escape of planetesimals. We suggest that the
  dissipation of the disk is related to the heavy bombardment phase in
  our Solar System.  Whether \textit{all} stars arrive on the main
  sequence surrounded by a disk cannot be established: some very young
  stars do not have a disk. And not all stars destroy their disk in a
  similar way: some stars as old as the Sun still have significant
  disks.
  
\end{sloppypar}

\end{abstract}

\section{Introduction}
In 1983, while using standard stars to calibrate the IRAS photometry,
\citet{auma:84} discovered that Vega ($\alpha$ Lyr), one of the best
calibrated and most used photometric standards in the visual
wavelength range, emits much more energy at mid- and far-infrared
wavelengths than its photosphere produces.  Because the star is not
reddened Aumann et al. proposed that the excess IR radiation is
emitted by small, interplanetary-dust particles in a disk rather
than in a spherical envelope.  This proposal was confirmed by
\citet{smit:84} who detected a flat source of scattered light around
$\beta$ Pic, one of the other Vega--like stars detected in the IRAS
data \citep{gill:86}, and the one with the strongest excess. The disk
around Vega and other main-sequence stars is the remnant of a much
stronger disk built up during the formation of the stars.
\citet{auma:84} pointed out that such disks have a lifetime much
shorter than the stellar age and therefore need to be rebuilt
continuously; collisions between asteroids are a probable source of
new dust \citep{weis:84}.  Except for the somewhat exceptional
case of $\beta$ Pic \citep{hobb:85} and in spite of several deep
searches no trace of any gas has ever been found in the disks around
main-sequence stars; see e.g. \citet{lise:99}.

Since 1984 the search for and the study of remnant disks has made
substantial progress by the discovery of numerous ``fatter'' disks
around pre-main sequence stars that contain dust and gas; for
overviews see \citet{beck:96}, \citet{sarg:93}, \citet{vand:98}. The
IRAS data base also contains detections of Vega-like disks around red
giant stars that have developed from A and F-type main-sequence stars
\citep{plet:97}.

The discovery of Aumann et al. has prompted deeper searches in the
IRAS data base with different strategies \citep{auma:85, walk:88,
  mann:98}. For a review see \citet{back:93}.  Recently
\citet{plet:99} have discussed these earlier results and concluded
that a significant excess at $60\mu$m is found in $13\pm 10\%$ of all
main sequence stars with spectral type A, F, G and K.  Unfortunately
all these studies based on IRAS data only were affected by severe
selection effects and did not answer important questions such as: Will
a star loose its disk when it grows older? On what time-scale?  Does
the presence of planets depend on the stellar main-sequence mass?  Do
multiple stars have disks more, or less frequently? Do stars that
formed in clusters have disks less often?  With such questions
unanswered we clearly do not understand the systematics of the
formation of solar systems.

Here we present results of a continuation with ISO \citep{kess:96} of
the succesful search of IRAS. Our aim has been to obtain a better
defined sample of stars. The major step forward in this paper is not
in the detection of more remnant disks, but in reliable information
about the presence \textit{or absence} of a disk.  Earlier reports on
results from our program have been given in \citet{habi:96},
\citet{domi:98}, \citet{jour:99} and \citet{habi:99}.

\section{Selecting and preparing the sample}

Stars were selected so that their photospheric flux was within our
sensitivity limit. Any excess would then appear immediately. We also
wanted to make certain that any excess flux should be attributed to a
circumstellar disk and not to some other property of the star, such as
circumstellar matter ejected during the stellar evolution or to the
presence of a red companion.

In selecting our stars we used the following criteria: 
\begin{itemize}
  
\item We selected main-sequence stars with an expected photospheric
  flux at 60\,\um\ larger than 30\, mJy. We started from a list of
  infrared flux densities calculated by \citet{john:83} for 2000 stars
  contained in a catalogue of stars within 25\, pc by \citet{wool:70}.
  The limit of 30\, mJy was based on the sensitivity of ISOPHOT
  \citep{lemk:96} as announced before launch by \citet{klaa:94}.
  
\item We removed all stars with peculiarities in their spectra for
  which an accurate infrared flux density could not be predicted. This
  made us eliminate all O and B stars (emission lines, free-free IR
  excess) and all M stars (molecular spectra not well understood).
  
\item We also removed all spectroscopic double stars. Visual double
  stars were rejected when the companion lies within 1 arcminute
  \textit{and} its $V$-magnitude differs by less than 5 magnitudes;
  any remaining companion will contribute less than 10\% of the
  60\,\um\  flux as one may show through Eq. \ref{eq:plets}.

\item We also excluded variable stars; the variability of all stars
has been checked \textit{a posteriori} using the photometry in the
band at 0.6\um\, given in the Hipparcos catalogue \citep{perr:97}; in
all cases the stellar magnitude is constant within 0.07 magnitude.

\end{itemize}

To illustrate what stars are bright enough to be included we use an
equation that gives the stellar colour, ($V-[60 \mu m]$), as a
function of $(B-V)$. The equation has been derived empirically from
IRAS data by \citet{wate:87}; we use a slightly different version
given by H. Plets (private communication):

\begin{eqnarray}
\label{eq:plets}
V-[60\mu m]& = &0.01+2.99(B-V)\nonumber\\
&&-1.06(B-V)^2+0.47(B-V)^3
\end{eqnarray}
The zero point in this equation has a formal error of 0.01.
Intrinsic, reddening-free $(B-V)$ values must be used, but all our
stars are nearby and we assume that the measured values are
reddening-free. A posteriori we checked that we may safely ignore the
reddening produced by the disks that we detected; only in the case of
$\beta$ Pic is a small effect expected. Adopting a flux density of
1.19 Jy for $[60\mu m]=0$ we find apparent-magnitude limits and
distance limits of suitable main-sequence stars as summarized in Table
\ref{tab:distlimits}.  The distance limit varies strongly with
spectral type.

\begin{table}[h!]
\caption{\label{tab:distlimits} Apparent magnitude and distance from 
the Sun (in parsec) of main-sequence stars with a 60\,\um\  flux density 
of 30\, mJy.}
%\scriptsize
\begin{tabular}{|c|cccccccc|}
\hline
Sp.Type &A0 & A5 & F0 & F5 & G0 & G5 & K0 & K5 \\
\hline
 $V$(mag)&4.0 & 4.4 & 4.8 & 5.2 & 5.7 & 6.0 & 6.8 & 7.4 \\
 $d$(pc)&45 & 31 & 25 & 19 & 15 & 13 & 10&7.5\\
\hline
\end{tabular}
\end{table}

\begin{table*}
\caption{\label{tab:stars} The stars of the sample}
\scriptsize
\begin{tabular}{|rrlrrrcrr|}
\hline
HD & HIP & Name & V & B-V &  d & Spect. & age & $T_\mathrm{eff}$  \\
   &     &      & mag & mag & pc&        & Gyrs& K                \\
(1)&(2)&(3)&(4)&(5)&(6)&(7)&(8)&(9)\\
\hline
$693$ & $910$ &  6 Cet              & $   4.89$ & $   0.49$ & $    18.9$ & F5V       & $5.13$ & $6210$\\
$1581$ & $1599$ &  $\zeta$ Tuc      & $   4.23$ & $   0.58$ & $     8.6$ & F9V       & $6.46$ & $5990$\\
$2151$ & $2021$ &  $\beta$ Hyi      & $   2.82$ & $   0.62$ & $     7.5$ & G2IV      & $5.37$ & $5850$\\
$4628$ & $3765$ &                   & $   5.74$ & $   0.89$ & $     7.5$ & K2V       & $7.94$ & $5050$\\
$4813$ & $3909$ &  $\phi^2$ Cet     & $   5.17$ & $   0.51$ & $    15.5$ & F7IV-V    & $1.38$ & $6250$\\
$7570$ & $5862$ &  $\nu$ Phe        & $   4.97$ & $   0.57$ & $    15.1$ & F8V       & $3.16$ & $6080$\\
$9826$ & $7513$ & $50$ And          & $   4.10$ & $   0.54$ & $    13.5$ & F8V       & $2.88$ & $6210$\\
$10700$ & $8102$ &  $\tau$ Cet      & $   3.49$ & $   0.73$ & $     3.6$ & G8V       & $7.24$ & $5480$\\
$10780$ & $8362$ &                  & $   5.63$ & $   0.80$ & $    10.0$ & K0V       & $2.82$ & $5420$\\
$12311$ & $9236$ &  $\alpha$ Hyi    & $   2.86$ & $   0.29$ & $    21.9$ & F0V       & $0.81$ & $7080$\\
$13445$ & $10138$ &                 & $   6.12$ & $   0.81$ & $    10.9$ & K0V       & $5.37$ & $5400$\\
$14412$ & $10798$ &                 & $   6.33$ & $   0.72$ & $    12.7$ & G8V       & $7.24$ & $5420$\\
$14802$ & $11072$ &  $\kappa$ For   & $   5.19$ & $   0.61$ & $    21.9$ & G2V       & $5.37$ & $5850$\\
$15008$ & $11001$ &  $\delta$ Hyi   & $   4.08$ & $   0.03$ & $    41.5$ & A3V       & $0.45$ & $8920$\\
$17051$ & $12653$ &  $\iota$ Hor    & $   5.40$ & $   0.56$ & $    17.2$ & G3IV      & $3.09$ & $6080$\\
$17925$ & $13402$ &                 & $   6.05$ & $   0.86$ & $    10.4$ & K1V       & $0.08$ & $5000$\\
$19373$ & $14632$ &  $\iota$ Per    & $   4.05$ & $   0.60$ & $    10.5$ & G0V       & $3.39$ & $6040$\\
$20630$ & $15457$ &  $\kappa^1$ Cet & $   4.84$ & $   0.68$ & $     9.2$ & G5Vv      & $0.30$ & $5750$\\
$20766$ & $15330$ & $\zeta^1$ Ret   & $   5.53$ & $   0.64$ & $    12.1$ & G2V       & $4.79$ & $5750$\\
$20807$ & $15371$ & $\zeta^2$ Ret   & $   5.24$ & $   0.60$ & $    12.1$ & G1V       & $7.24$ & $5890$\\
$22001$ & $16245$ & $\kappa$ Ret    & $   4.71$ & $   0.41$ & $    21.4$ & F5IV-V    & $2.04$ & $6620$\\
$22049$ & $16537$ & $\epsilon$ Eri  & $   3.72$ & $   0.88$ & $     3.2$ & K2V       & $0.33$ & $5000$\\
$22484$ & $16852$ & $10$ Tau        & $   4.29$ & $   0.58$ & $    13.7$ & F9V       & $5.25$ & $5980$\\
$23249$ & $17378$ & $\delta$ Eri    & $   3.52$ & $   0.92$ & $     9.0$ & K2V       & $7.59$ & $5000$\\
$26965$ & $19849$ & $o^2$ Eri       & $   4.43$ & $   0.82$ & $     5.0$ & K1V       & $7.24$ & $5100$\\
$30495$ & $22263$ & $58$ Eri        & $   5.49$ & $   0.63$ & $    13.3$ & G3V       & $0.21$ & $5820$\\
$33262$ & $23693$ & $\zeta$ Dor     & $   4.71$ & $   0.53$ & $    11.7$ & F7V       & $2.95$ & $6160$\\
$34411$ & $24813$ & $\lambda$ Aur   & $   4.69$ & $   0.63$ & $    12.7$ & G0V       & $6.76$ & $5890$\\
$37394$ & $26779$ &                 & $   6.21$ & $   0.84$ & $    12.2$ & K1V       & $0.34$ & $5100$\\
$38392$ &         &                 & $   6.15$ & $   0.94$ & $     9.0$ & K2V       & $0.87$ & $4950$\\
$38393$ & $27072$ & $\gamma$ Lep    & $   3.59$ & $   0.48$ & $     9.0$ & F7V       & $1.66$ & $6400$\\
$38678$ & $27288$ & $\zeta$ Lep     & $   3.55$ & $   0.10$ & $    21.5$ & A2Vann    & $0.37$ & $8550$\\
$39060$ & $27321$ & $\beta$ Pic     & $   3.85$ & $   0.17$ & $    19.3$ & A3V       & $0.28$ & $8040$\\
$43834$ & $29271$ & $\alpha$ Men    & $   5.08$ & $   0.71$ & $    10.2$ & G5V       & $7.24$ & $5630$\\
$48915$ & $32349$ & $\alpha$ CMa    & $  -1.44$ & $   0.01$ & $     2.6$ & A0m       &        & $9920$\\
$50281$ & $32984$ &                 & $   6.58$ & $   1.07$ & $     8.7$ & K3V       & $2.63$ & $5000$\\
$61421$ & $37279$ & $\alpha$ CMi    & $   0.40$ & $   0.43$ & $     3.5$ & F5IV-V    & $1.70$ & $6700$\\
$74956$ & $42913$ & $\delta$ Vel    & $   1.93$ & $   0.04$ & $    25.0$ & A1V       & $0.35$ & $9200$\\
$75732$ & $43587$ & $\rho^1$ Cnc    & $   5.96$ & $   0.87$ & $    12.5$ & G8V       & $5.01$ & $5300$\\
$80007$ & $45238$ & $\beta$ Car     & $   1.67$ & $   0.07$ & $    34.1$ & A2IV      &        & $8600$\\
$95418$ & $53910$ & $\beta$ UMa     & $   2.34$ & $   0.03$ & $    24.4$ & A1V       & $0.36$ & $9530$\\
$102647$ & $57632$ & $\beta$ Leo    & $   2.14$ & $   0.09$ & $    11.1$ & A3Vvar    & $0.24$ & $8580$\\
$102870$ & $57757$ & $\beta$ Vir    & $   3.59$ & $   0.52$ & $    10.9$ & F8V       & $2.63$ & $6180$\\
$103287$ & $58001$ & $\gamma$ UMa   & $   2.41$ & $   0.04$ & $    25.7$ & A0V SB    & $0.38$ & $9440$\\
$106591$ & $59774$ & $\delta$ UMa   & $   3.32$ & $   0.08$ & $    25.0$ & A3Vvar    & $0.48$ & $8630$\\
$110833$ & $62145$ &                & $   7.01$ & $   0.94$ & $    15.1$ & K3V       & $12.60$ & $5000$\\
$112185$ & $62956$ & $\epsilon$ UMa & $   1.76$ & $  -0.02$ & $    24.8$ & A0p       & $0.30$ & $9780$\\
$114710$ & $64394$ & $\beta$ Com    & $   4.23$ & $   0.57$ & $     9.2$ & G0V       & $3.63$ & $6030$\\
$116842$ & $65477$ & $80$ UMa       & $   3.99$ & $   0.17$ & $    24.9$ & A5V       & $0.32$ & $8000$\\
$126660$ & $70497$ & $\theta$ Boo   & $   4.04$ & $   0.50$ & $    14.6$ & F7V       & $2.95$ & $6280$\\
$128167$ & $71284$ & $\sigma$ Boo   & $   4.46$ & $   0.36$ & $    15.5$ & F3Vwvar   & $1.70$ & $6770$\\
$134083$ & $73996$ & $45$ Boo       & $   4.93$ & $   0.43$ & $    19.7$ & F5V       & $1.82$ & $6500$\\
$139664$ & $76829$ & g Lup          & $   4.64$ & $   0.41$ & $    17.5$ & F5IV-V    & $1.12$ & $6680$\\
$142373$ & $77760$ & $\chi$ Her     & $   4.60$ & $   0.56$ & $    15.9$ & F9V       & $8.51$ & $5840$\\
$142860$ & $78072$ & $\gamma$ Ser   & $   3.85$ & $   0.48$ & $    11.1$ & F6V       & $3.24$ & $6330$\\
$149661$ & $81300$ & $12$ Oph       & $   5.77$ & $   0.83$ & $     9.8$ & K2V       & $2.09$ & $5200$\\
$154088$ & $83541$ &                & $   6.59$ & $   0.81$ & $    18.1$ & K1V       & $7.24$ & $5000$\\
$156026$ & $84478$ &                & $   6.33$ & $   1.14$ & $     6.0$ & K5V       & $0.63$ & $4350$\\
$157214$ & $84862$ & $72$ Her       & $   5.38$ & $   0.62$ & $    14.4$ & G0V       & $7.24$ & $5790$\\
$157881$ & $85295$ &                & $   7.54$ & $   1.36$ & $     7.7$ & K7V       & $5.25$ & $3950$\\
$160691$ & $86796$ & $\mu$ Ara      & $   5.12$ & $   0.69$ & $    15.3$ & G5V       & $6.17$ & $5750$\\
$161797$ & $86974$ & $\mu$ Her      & $   3.42$ & $   0.75$ & $     8.4$ & G5IV      & $4.79$ & $5670$\\
$166620$ & $88972$ &                & $   6.38$ & $   0.88$ & $    11.1$ & K2V       & $7.24$ & $4970$\\
$172167$ & $91262$ & $\alpha$ Lyr   & $   0.03$ & $   0.00$ & $     7.8$ & A0Vvar    & $0.35$ & $9620$\\
$173667$ & $92043$ & $110$ Her      & $   4.19$ & $   0.48$ & $    19.1$ & F6V       & $2.40$ & $6370$\\
$185144$ & $96100$ & $\sigma$ Dra   & $   4.67$ & $   0.79$ & $     5.8$ & K0V       & $5.50$ & $5330$\\
$185395$ & $96441$ & $\theta$ Cyg   & $   4.49$ & $   0.40$ & $    18.6$ & F4V       & $1.29$ & $6750$\\
$187642$ & $97649$ & $\alpha$ Aql   & $   0.76$ & $   0.22$ & $     5.1$ & A7IV-V    & $1.23$ & $7550$\\
$188512$ & $98036$ & $\beta$ Aql    & $   3.71$ & $   0.86$ & $    13.7$ & G8IV      & $4.27$ & $5500$\\
$190248$ & $99240$ & $\delta$ Pav   & $   3.55$ & $   0.75$ & $     6.1$ & G5IV-Vvar & $5.25$ & $5650$\\
$191408$ & $99461$ &                & $   5.32$ & $   0.87$ & $     6.0$ & K2V       & $7.24$ & $4700$\\
$192310$ & $99825$ &                & $   5.73$ & $   0.88$ & $     8.8$ & K3V       &        & $5000$\\
$197692$ & $102485$ & $\psi$ Cap    & $   4.13$ & $   0.43$ & $    14.7$ & F5V       & $2.00$ & $6540$\\
$198149$ & $102422$ & $\eta$ Cep    & $   3.41$ & $   0.91$ & $    14.3$ & K0IV      & $7.94$ & $5000$\\
$203280$ & $105199$ & $\alpha$ Cep  & $   2.45$ & $   0.26$ & $    15.0$ & A7IV-V    & $0.89$ & $7570$\\
$203608$ & $105858$ & $\gamma$ Pav  & $   4.21$ & $   0.49$ & $     9.2$ & F6V       & $10.50$ & $6150$\\
$207129$ & $107649$ &               & $   5.57$ & $   0.60$ & $    15.6$ & G2V       & $6.03$ & $5930$\\
$209100$ & $108870$ & $\zeta$ Ind   & $   4.69$ & $   1.06$ & $     3.6$ & K5V       & $1.29$ & $4600$\\
$215789$ & $112623$ & $\epsilon$ Gru & $  3.49$ & $   0.08$ & $    39.8$ & A3V       & $0.54$ & $8420$\\
$216956$ & $113368$ & $\alpha$ Psa  & $   1.17$ & $   0.15$ & $     7.7$ & A3V       & $0.22$ & $8680$\\
$217014$ & $113357$ & $51$ Peg      & $   5.45$ & $   0.67$ & $    15.4$ & G5V       & $5.13$ & $5810$\\
$219134$ & $114622$ &               & $   5.57$ & $   1.00$ & $     6.5$ & K3Vvar    & $12.60$ & $4800$\\
$222368$ & $116771$ & $\iota$ Psc   & $   4.13$ & $   0.51$ & $    13.8$ & F7V       & $3.80$ & $6190$\\
$222404$ & $116727$ & $\gamma$ Cep  & $   3.21$ & $   1.03$ & $    13.8$ & K1IV      & $8.91$ & $5000$\\
\hline
\end{tabular}
\end{table*}

Table \ref{tab:stars} contains basic data on all stars from the
sample for which we present ISO data. Columns 1 and 2 contain the
number of the star in the HD and in the Hipparcos Catalogue
\citep{perr:97} and column 3 the name. $V$ and $B-V$ have been taken
from the Geneva photometric catalogue \citep{kunz:97}.  Columns 6 and
7 contain the distance and the spectral type as given in the Hipparcos
Catalog \citep{perr:97}. The age given in column 8 is from
\citet{lach:99}, where errors in the age determinations are discussed.
The effective temperature in column 11 has been derived by fitting
Kurucz' model atmospheres to the Geneva photometry; we will need this
temperature to calculate the dust mass from the flux-density excess at
60\,\um.

\section{Measurements, data reduction, checks}

\subsection{Measurements}
Pre-launch recommendations made us start with chopped measurements
(observing mode PHT03; see \citet{laur:00}) at 60, 90, 135 and 170
$\mu$m.  After a few months of operation of the satellite it appeared
that at 60 and 90 $\mu$m the on-off signal was strongly distorted by
transients in the responsitivity of the detectors. Similarly, chopping
appeared to be an inadequate observing mode at 135 and at 170 $\mu$m
because of confusion with structure in the background from infrared
cirrus. We therefore switched to the observing mode PHT22 and made
minimaps.  Minimaps consumed more observing time and we therefore
dropped the observations at 90 and 135 $\mu$m. We tried to reobserve
in minimap mode those targets that had already been observed in
chopped mode (using extra time allocated when ISO lived longer than
expected) but succeeded only partially: several targets had left the
observing window. In total we used 65 hrs of observations. In this
article we discuss only the stellar flux densities derived from the 60
and 170\,$\mu$m minimaps.  Appendix A contains a detailed description
of our measurement procedure.

Instrumental problems (mainly detector memory effects) made us
postpone the reduction of the chopped measurements until a later date;
this applies also to the many (all chopped) measurements at 25\,\um. 

We added published \citep{abra:98} ISOPHOT measurements of five A-type
stars ($\beta$ UMa, $\gamma$ UMa, $\delta$ UMa, $\epsilon$ UMa and 80
UMa). The measurements have been obtained in a different mode from our
observations, but we treat all measurements equally. These five stars
are all at about 25\, pc \citep{perr:97}, sufficiently nearby to allow
detection of the photospheric flux. These stars are spectroscopic
doubles and they do not fulfill all of our selection criteria; below
we argue why we included them anyhow.  \citet{abra:98} present ISOPHOT
measurements of four more stars, which they assume to be at the same
distance because all nine stars are supposed to be members of an
equidistant group called the ``Ursa Major stream''. The Hipparcos
measurements \citep{perr:97}, however, show that four of the nine
stars are at a distance of 66\, pc and thus too far away to be
useful for our purposes.

\subsection{Data reduction}

All our data have been reduced using standard calibration tables and
the processing steps of OLP6/PIA7. These steps include the
instrumental corrections and photometric calibration of the data.  At
the time when we reduced our data there did not yet exist a standard
procedure to extract the flux. We therefore developed and used our own
method- see appendix B.

Later versions of the software which contain upgrades of the
photometric calibration do not significantly alter our photometric
results and the conclusions of this paper remain unchanged. For each
filter the observing mode gave two internal calibration measurements
which were closely tuned to the actual sky brightness. This makes the
absolute calibration insensitive to instrumental effects as
filter-to-filter calibrations and signal non-linearities which were
among others the main photometric calibration improvements for the
upgrades. In addition, it is standard procedure to ensure that each
upgrade does not degrade the photometric calibration of the validated
modes of the previous processing version.

\begin{table*}

\caption{\normalsize\label{tab:60um} 60\um{} data:
see text for an explanation of the various columns}
\scriptsize
\begin{center}
\begin{tabular}{|rr|rr|r|rr|rr|l|} 
\hline
HD                        &   % 1
ISO\_id                   &   % 2     TDT  
$F_{\nu}$                 &   % 3 or 6 or 5; add * for IRAS FSC, ** for PSC  
                              %  psf and colour-corrected (raw flux /0.69/1.06)
$\sigma_{\nu}$            &   % 4
$F_{\nu}^{\mathrm{pred}}$ &   % 5  
$F_{\nu}^{\mathrm{exc}}$ &    % 6
% $\frac{\displaystyle F_{\nu}^{\mathrm{exc}}}{\displaystyle \sigma_{\nu}^{\mathrm{RJ}}}$ & % 7
$F_{\nu}^{\mathrm{exc}}$/$\sigma_{\nu}$  &  %  7 (col6 / col4) 
$F_{\nu}^{\mathrm{disk}}$ &   % 8     de-colour-corrected    (* 1.06)
$log\, \tau^{\mathrm{disk}}_{60}$         &     % 9
Reference \\     % 10
 & & mJy & mJy & mJy  & mJy &   & mJy & & \\
(1)&(2)&(3)&(4)&(5)&(6)&(7)&(8)&(9)&(10)\\
\hline
$693$   & $74900501$ & $33$ & $34$ & $43$ & $-10$ & $<1$ & &$<-4.6$ & ISO minimap \\
$1581$  & $69700102$ & $94$ & $19$ & $94$ & $0$ & $<1$ & &$<-5.0$  & ISO minimap \\
$2151$  &            & $351$ & $100$ & $376$ & $-25$ & $<1$ & & & IFSC \\
$4628$  & $61901104$ & $41$ & $24$ & $44$ & $-4$ & $<1$ & &$<-4.1$ & ISO minimap \\
$4813$  & $61901705$ & $55$ & $37$ & $34$ & $20$ & $<1$ & &$<-4.3$ & ISO minimap \\
$7570$  & $72002506$ & $85$ & $25$ & $47$ & $38$ & $1.6$ & &$<-4.4$ & ISO minimap \\
$9826$  & $61503786$ & $100$ & $22$ & $98$ & $2$ & $<1$ & &$<-5.0$ & ISO minimap \\
$10700$ & $75701121$ & $433$ & $37$ & $253$ & $180$ & $4.9$ & $190$ & $-4.6$ & ISO minimap \\ %yes
$10780$ & $61503507$ & $43$ & $30$ & $41$ & $3$ & $<1$ & & $<-4.1$& ISO minimap \\
$12311$ & $69100108$ & $189$ & $18$ & $178$ & $10$ & $<1$ & &$<-5.6$ & ISO minimap \\
$13445$ & $73100771$ & $25$ & $24$ & $26$ & $-2$ & $<1$ & &$<-4.1$ & ISO minimap \\
$14412$ & $76301673$ & $-12$ & $14$ & $18$ & $-30$ & $<1$ & &$<-4.4$ & ISO minimap \\
$14802$ & $80201174$ & $56$ & $19$ & $41$ & $14$ & $<1$ & &$<-4.5$ & ISO minimap \\
$15008$ & $71801511$ & $41$ & $9$ & $30$ & $10$ & $1.1$ & &$<-5.4$ & ISO minimap \\
$17051$ & $76500413$ & $45$ & $15$ & $31$ & $14$ & $<1$ & &$<-4.5$ & ISO minimap \\
$17925$ & $78100314$ & $104$ & $24$ & $31$ & $73$ & $3.1$ & $80$ & $-3.9$ & ISO minimap \\ %yes
$19373$ & $81001847$ & $122$ & $13$ & $116$ & $6$ & $<1$ & &$<-5.2$ & ISO minimap \\
$20630$ & $79201553$ & $9$ & $33$ & $66$ & $-57$ & $<1$ & &$<-4.9$ & ISO minimap \\
$20766$ & $69100715$ & $34$ & $24$ & $32$ & $2$ & $<1$ & &$<-4.4$ & ISO minimap \\
$20807$ & $57801756$ & $30$ & $13$ & $39$ & $-9$ & $<1$ & &$<-4.9$ & ISO minimap \\
$22001$ & $69100659$ & $52$ & $9$ & $42$ & $9$ & $1.0$ & &$<-5.1$ & ISO minimap \\
$22049$ &            & $1250$ & $100$ & $278$ & $967$ & $9.7$ & $1260$ & & IFSC \\%yes
$22484$ & $79501562$ & $141$ & $27$ & $89$ & $51$ & $1.9$ & &$<-4.6$ & ISO minimap \\
$23249$ &            & $270$ & $100$ & $363$ & $-92$ & $<1$ & & & IFSC \\
$26965$ & $84801865$ & $121$ & $21$ & $128$ & $-7$ & $<1$ & &$<-4.7$ & ISO minimap \\
$30495$ & $83901668$ & $174$ & $31$ & $33$ & $141$ & $4.5$ & $150$ & $-4.1$ & ISO minimap\\ %yes
$33262$ & $58900871$ & $81$ & $21$ & $55$ & $26$ & $1.3$ & &$<-4.7$ & ISO minimap \\
$34411$ & $83801474$ & $63$ & $13$ & $69$ & $-5$ & $<1$ & &$<-5.0$ & ISO minimap \\
$37394$ & $83801977$ & $40$ & $13$ & $26$ & $14$ & $1.0$ & &$<-4.0$ & ISO minimap \\
$38392$ & $70201402$ & $31$ & $24$ & $34$ & $-2$ & $<1$ & &$<-3.9$ & ISO minimap \\
$38393$ & $70201305$ & $160$ & $25$ & $138$ & $22$ & $<1$ & &$<-5.1$ & ISO minimap\\
$38678$ & $69202308$ & $349$ & $22$ & $60$ & $289$ & $13.3$ & $310$ & $-4.7$ & ISO minimap \\ %yes
$39060$ & $70201080$ & $14700$ & $346$ & $54$ & $14650$ & $42.4$ & $15500$ & $-2.8$ & ISO minimap \\ %yes
$43834$ & $62003217$ & $48$ & $16$ & $56$ & $-8$ & $<1$ & &$<-4.8$ & ISO minimap \\
$48915$ & $72301711$ & $4230$ & $155$ & $4650$ & $-420$ & $<1$ & &$<-7.5$ & ISO minimap \\
$50281$ & $71802114$ & $-4$ & $16$ & $30$ & $-33$ & $<1$ & &$<-4.3$ & ISO minimap \\
$61421$ &            & $2290$ & $100$ & $2350$ & $-59$ & $<1$ & &$<-6.0$ & IFSC \\ 
$74956$ &            & $399$ & $100$ & $226$ & $173$ & $1.7$ & & & IPSC \\
$75732$ & $17800102$ & $160$ & $28$ & $35$ & $126$ & $4.4$ & $130$ & $-3.8$ & Dominik et al. 1998 \\ %yes
$80007$ &            & $284$ & $100$ & $311$ & $-27$ & $<1$ & & & IFSC \\
$95418$ & $19700563$ & $539$ & $135$ & $152$ & $387$ & $2.9$ & $410$ & $-5.0$ &  \'Abrah\'am et al. 1998 \\ %yes
$102647$ &           & $784$ & $100$ & $213$ & $571$ & $5.7$ & $750$ & $-4.8$ & IFSC\\ %yes
$102870$ &           & $137$ & $100$ & $150$ & $-14$ & $<1$ & &$<-4.6$ & IFSC\\
$103287$ & $19500468$ & $164$ & $41$ & $147$ & $17$ & $<1$ & &$<-5.5$ & \'Abrah\'am et al. 1998 \\
$106591$ & $19700973$ & $94$ & $59$ & $69$ & $25$ & $<1$ & &$<-4.9$ & \'Abrah\'am et al. 1998 \\
$110833$ & $60000526$ & $-7$ & $14$ & $15$ & $-22$ & $<1$ & & & ISO minimap \\
$112185$ & $34600578$ & $322$ & $81$ & $223$ & $99$ & $1.2$ & &$<-5.4$ & \'Abrah\'am et al. 1998 \\
$114710$ & $61000119$ & $106$ & $34$ & $93$ & $13$ & $<1$ & &$<-4.7$ & ISO minimap \\
$116842$ & $19500983$ & $40$ & $34$ & $47$ & $-7$ & $<1$ & & & \'Abrah\'am et al. 1998 \\
$126660$ & $61000834$ & $93$ & $21$ & $95$ & $-2$ & $<1$ & &$<-5.1$ & ISO minimap \\
$128167$ & $61001236$ & $100$ & $19$ & $48$ & $52$ & $2.8$ & $55$ & $-5.0$ & ISO minimap \\ %yes
$134083$ & $61001337$ & $70$ & $26$ & $36$ & $34$ & $1.3$ & &$<-4.5$ & ISO minimap \\
$139664$ & $64700880$ & $488$ & $48$ & $45$ & $442$ & $9.2$ & $470$ & $-4.0$ & ISO minimap \\ %yes
$142373$ & $61001139$ & $71$ & $21$ & $64$ & $6$ & $<1$ & &$<-4.8$ & ISO minimap \\
$142860$ & $63102981$ & $113$ & $24$ & $109$ & $4$ & $<1$ & &$<-5.1$ & ISO minimap \\
$149661$ & $80700365$ & $56$ & $22$ & $38$ & $18$ & $<1$ & &$<-4.1$ & ISO minimap \\
$154088$ & $64702041$ & $40$ & $55$ & $17$ & $22$ & $<1$ & &$<-3.3$ & ISO minimap \\
$156026$ & $64702142$ & $30$ & $34$ & $43$ & $-13$ & $<1$ & &$<-3.4$ & ISO minimap \\
$157214$ & $71000144$ & $27$ & $26$ & $36$ & $-8$ & $<1$ & &$<-4.4$ & ISO minimap \\
$157881$ & $65000845$ & $41$ & $22$ & $24$ & $17$ & $<1$ & &$<-2.6$ & ISO minimap \\
$160691$ & $64402347$ & $73$ & $18$ & $52$ & $21$ & $1.2$ & &$<-4.5$ & ISO minimap \\
$161797$ &            & $222$ & $100$ & $281$ & $-59$ & $<1$ & &$<-4.6$ & IFSC\\
$166620$ & $71500648$ & $41$ & $21$ & $24$ & $17$ & $<1$ & &$<-3.8$ & ISO minimap \\
$172167$ & $71500582$ & $6530$ & $217$ & $1170$ & $5360$ & $24.7$ & $5700$ & $-4.8$ & ISO minimap \\ %yes
$173667$ & $71500883$ & $78$ & $12$ & $80$ & $-1$ & $<1$ & &$<-5.3$ & ISO minimap \\
$185144$ & $69500449$ & $92$ & $21$ & $96$ & $-5$ & $<1$ & &$<-4.7$ & ISO minimap \\
$185395$ & $69301251$ & $63$ & $34$ & $51$ & $12$ & $<1$ & &$<-4.7$ & ISO minimap \\
$187642$ & $72400584$ & $1010$ & $66$ & $1050$ & $-41$ & $<1$ & &$<-6.1$ & ISO minimap \\
$188512$ &            & $257$ & $100$ & $269$ & $-12$ & $<1$ & & & IFSC \\
$190248$ &            & $174$ & $100$ & $250$ & $-76$ & $<1$ & &$<-4.6$ & IFSC\\
$191408$ & $72501252$ & $43$ & $37$ & $62$ & $-19$ & $<1$ & &$<-4.0$ & ISO minimap \\
$192310$ & $70603454$ & $73$ & $13$ & $44$ & $29$ & $2.2$ & &$<-4.1$ & ISO minimap \\
$197692$ & $70603356$ & $68$ & $18$ & $76$ & $-8$ & $<1$ & &$<-5.2$ & ISO minimap \\
$198149$ &            & $552$ & $100$ & $393$ & $159$ & $1.6$ & & & IFSC \\
$203280$ & $61002158$ & $253$ & $49$ & $243$ & $10$ & $<1$ & &$<-5.4$ & ISO minimap \\
$203608$ & $72300260$ & $109$ & $21$ & $80$ & $30$ & $1.4$ & &$<-4.9$ & ISO minimap \\
$207129$ & $13500820$ & $275$ & $55$ & $29$ & $246$ & $4.5$ & $260$ & $-3.8$ & Jourdain de Muizon et al. 1999 \\ %yes
$209100$ & $70800865$ & $146$ & $24$ & $166$ & $-19$ & $<1$ & &$<-4.4$ & ISO minimap \\
$215789$ & $71801167$ & $74$ & $16$ & $60$ & $14$ & $<1$ & &$<-5.4$ & ISO minimap \\
$216956$ & $71800269$ & $6930$ & $204$ & $605$ & $6320$ & $31.0$ & $6700$ & $-4.3$ & ISO minimap \\ %yes
$217014$ & $73601191$ & $1$ & $24$ & $37$ & $-36$ & $<1$ & &$<-4.7$ & ISO minimap \\
$219134$ & $75100962$ & $17$ & $15$ & $65$ & $-48$ & $<1$ & & & ISO minimap \\
$222368$ & $74702964$ & $83$ & $18$ & $90$ & $-7$ & $<1$ & &$<-5.2$ & ISO minimap \\
$222404$ &            & $537$ & $100$ & $607$ & $-70$ & $<1$ & & & IFSC \\
\hline
\end{tabular}
\end{center}
\end{table*}

\begin{table*}
\caption{\normalsize\label{tab:170um}  Flux densities 
measured at 170\um{}. The various columns have 
the same meaning as in the previous table }
\scriptsize
\begin{center}
\begin{tabular}{|rr|rr|r|rr|rr|} 
\hline
HD                        &  % 1
ISO\_id                   &  % 2
$F_{\nu}$                 &  % 3   psf and colour corrected (raw flux/0.64/1.2)
$\sigma_{\nu}$            &       % 4   psf and colour corrected
$F_{\nu}^{\mathrm{pred}}$ &       % 5
$F_{\nu}^{\mathrm{exc}}$ &        % 6   colour corrected  (col3 -col5)
% $\frac{\displaystyle F_{\nu}^{\mathrm{exc}}}{\displaystyle 
% \sigma_{\nu}}$ & % 7   (col6 / col4)
$F_{\nu}^{\mathrm{exc}}$/$\sigma_{\nu}$  &  %  7 (col6 / col4) 
$F_{\nu}^{\mathrm{disk}}$ &      % 8    de-colour corrected    (col6  *1.2)
$log\, \tau^{\mathrm{disk}}_{170}$\\                                                %9
& & mJy  & mJy & mJy  & mJy  & & mJy & \\
(1)&(2)&(3)&(4)&(5)&(6)&(7)&(8)&(9)\\
\hline
$   693$ & $ 37500903$ & $   12$ & $   13$ & $   5$ & $    7$ & $  <1 $ &&\\
$  4628$ & $ 39502509$ & $  -10$ & $    9$ & $   5$ & $  -15$ & $  <1 $ &&\\
$  4813$ & $ 38701512$ & $   12$ & $   10$ & $   4$ & $    8$ & $  <1 $ &&\\
$  7570$ & $ 38603615$ & $   14$ & $   36$ & $   5$ & $    9$ & $  <1 $ &&\\
$  9826$ & $ 42301521$ & $  -22$ & $   82$ & $  10$ & $  -32$ & $  <1 $ &&\\
$ 10700$ & $ 39301218$ & $  125$ & $   21$ & $  28$ & $   97$ & $ 4.7 $ & $  120$ & $ -5.4$ \\
$ 10780$ & $ 45701321$ & $  180$ & $  626$ & $   4$ & $  176$ & $  <1 $ &&\\
$ 12311$ & $ 69100209$ & $   98$ & $   52$ & $  19$ & $   79$ & $ 1.5 $ &&\\
$ 14412$ & $ 40101733$ & $  -20$ & $   12$ & $   2$ & $  -22$ & $  <1 $ &&\\
$ 14802$ & $ 40301536$ & $    1$ & $   16$ & $   5$ & $   -4$ & $  <1 $ &&\\
$ 15008$ & $ 75000612$ & $   23$ & $   49$ & $   3$ & $   20$ & $  <1 $ &&\\
$ 17051$ & $ 41102842$ & $    4$ & $    9$ & $   3$ & $    1$ & $  <1 $ &&\\
$ 19373$ & $ 81001848$ & $ -163$ & $  182$ & $  11$ & $ -174$ & $  <1 $ &&\\
$ 20630$ & $ 79201554$ & $ -122$ & $   85$ & $   6$ & $ -128$ & $  <1 $ &&\\
$ 20807$ & $ 57801757$ & $   73$ & $   17$ & $   4$ & $   69$ & $ 4.1 $ & $   80$ & $ -4.8$ \\
$ 22001$ & $ 69100660$ & $  -49$ & $   25$ & $   5$ & $  -54$ & $  <1 $ &&\\
$ 22484$ & $ 79501563$ & $    7$ & $   21$ & $   9$ & $   -2$ & $  <1 $ &&\\
$ 26965$ & $ 84801866$ & $   60$ & $   38$ & $  16$ & $   44$ & $ 1.2 $ &&\\
$ 30495$ & $ 83901669$ & $   51$ & $   25$ & $   4$ & $   47$ & $ 1.9 $ &&\\
$ 33262$ & $ 58900872$ & $  -33$ & $   33$ & $   6$ & $  -39$ & $  <1 $ &&\\
$ 34411$ & $ 83801475$ & $  -59$ & $   98$ & $   7$ & $  -66$ & $  <1 $ &&\\
$ 37394$ & $ 83801978$ & $   61$ & $   57$ & $   3$ & $   58$ & $ 1.0 $ &&\\
$ 38392$ & $ 70201403$ & $   25$ & $   20$ & $   4$ & $   21$ & $ 1.1 $ &&\\
$ 38393$ & $ 70201306$ & $   68$ & $    8$ & $  14$ & $   54$ & $ 6.9 $ & $  65$ & $ -5.4$ \\
$ 38678$ & $ 69202309$ & $   22$ & $   48$ & $   6$ & $   16$ & $  <1 $ &&\\
$ 39060$ & $ 70201081$ & $ 3807$ & $  143$ & $   6$ & $ 3801$ & $26.5 $ & $ 4600$ & $ -3.2$ \\
$ 48915$ & $ 72301712$ & $  184$ & $  401$ & $ 456$ & $ -272$ & $  <1 $ &&\\
$ 50281$ & $ 71802115$ & $ -826$ & $  268$ & $   2$ & $ -828$ & $  <1 $ &&\\
$ 95418$ & $ 19700564$ & $  133$ & $   73$ & $  13$ & $  120$ & $ 1.6 $ &&\\
$ 103287$ & $ 19500469$ & $   95$ & $  117$ & $  12$ & $   83$ & $  <1 $ &&\\
$ 106591$ & $ 33700130$ & $   -5$ & $   17$ & $   7$ & $  -12$ & $  <1 $ &&\\
$ 110833$ & $ 60000527$ & $  -31$ & $   23$ & $    $ & $  -31$ & $  <1 $ &&\\
$ 112185$ & $ 34600579$ & $  -35$ & $   65$ & $  21$ & $  -56$ & $  <1 $ &&\\
$ 126660$ & $ 61000935$ & $  -39$ & $   20$ & $  10$ & $  -49$ & $  <1 $ &&\\
$ 128167$ & $ 39400840$ & $   56$ & $   12$ & $   6$ & $   50$ & $ 4.3 $ & $  60$ & $ -5.0$ \\
$ 139664$ & $ 29101241$ & $  122$ & $  207$ & $   5$ & $  117$ & $  <1 $ &&\\
$ 142373$ & $ 62600340$ & $  -12$ & $   31$ & $   8$ & $  -20$ & $  <1 $ &&\\
$ 142860$ & $ 30300242$ & $   31$ & $   73$ & $  12$ & $   19$ & $  <1 $ &&\\
$ 149661$ & $ 30400943$ & $  113$ & $   69$ & $   4$ & $  109$ & $ 1.6 $ &&\\
$ 154088$ & $ 45801569$ & $ -138$ & $  137$ & $   2$ & $ -140$ & $  <1 $ &&\\
$ 156026$ & $ 83400343$ & $ -383$ & $  318$ & $   6$ & $ -389$ & $  <1 $ &&\\
$ 157214$ & $ 33600844$ & $  -31$ & $   34$ & $   4$ & $  -35$ & $  <1 $ &&\\
$ 160691$ & $ 29101345$ & $ -171$ & $   65$ & $   5$ & $ -176$ & $  <1 $ &&\\
$ 166620$ & $ 36901487$ & $   -9$ & $   21$ & $   3$ & $  -12$ & $  <1 $ &&\\
$ 172167$ & $ 44300846$ & $ 2621$ & $  142$ & $ 123$ & $ 2498$ & $17.6 $ & $ 3000$ & $  -4.8$ \\
$ 173667$ & $ 31902147$ & $  -53$ & $   91$ & $   8$ & $  -61$ & $  <1 $ &&\\
$ 185395$ & $ 35102048$ & $  -35$ & $   26$ & $   5$ & $  -40$ & $  <1 $ &&\\
$ 197692$ & $ 70603857$ & $   27$ & $   34$ & $   8$ & $   19$ & $  <1 $ &&\\
$ 203608$ & $ 72300361$ & $  -52$ & $    9$ & $  10$ & $  -62$ & $  <1 $ &&\\
$ 207129$ & $ 34402149$ & $  293$ & $   23$ & $   3$ & $  290$ & $12.4 $ & $  350$ & $ -4.0$ \\
$ 217014$ & $ 37401642$ & $  -57$ & $   23$ & $   4$ & $  -61$ & $  <1 $ &&\\
$ 222368$ & $ 37800836$ & $  -30$ & $   59$ & $  10$ & $  -40$ & $  <1 $ &&\\
\hline
\end{tabular}
\end{center}
\footnotesize
{{\it Note}: Fluxes in columns 3,4,6,7 are corrected for point spread function and Rayleigh-Jeans
colour-correction, i.e. the inband flux has been divided by 0.64 (psf) and by 1.2 (cc). 
The ``Excess'' in column 8 is ``de-colour-corrected'' from column 6.}
\normalsize
\end{table*}

\section{Results}

\subsection{Flux densities at 60$\mu$m}
The flux densities at 60\,\um\ are presented in Table \ref{tab:60um}.
The content of each column is as follows: (1): the HD number; (2): the
TDT number as used in the ISO archive; (3) the flux density corrected
for bandwidth effects (assuming that the spectrum is characterized by
the Rayleigh-Jeans equation) and for the fact that the stellar flux
extended over more than 1 pixel; (4) the error estimate assigned by
the ISOPHOT software to the flux measurement in column (3); for the
IRAS measurements the error has been put at 100\, mJy; (5) the flux
expected from the stellar photosphere, $F^{\mathrm{pred}}_{\nu}$ as
derived from Eq. \ref{eq:plets} using the $V$ and $(B-V)$ values in
Table \ref{tab:stars}; (6): the difference between columns (3) and
(5); we call it the ``excess flux'', $F_{\nu}^{\mathrm{exc}}$; (7):
the ratio of the excess flux compared to the measurement error given
in column (4); when we concluded that the excess is real and not a
measurement error we recalculated the monochromatic flux density by
assuming a flat spectrum within the ISOPHOT 60 $\mu$m bandwidth; the
result is in column (8) and is called $F_{\nu}^{\mathrm{disk}}$.
Column (9) shows an estimate of $\tau_{60}^{\mathrm{disk}}$, the
optical depth of the disk at visual wavelengths, but estimated from
the flux density at $60\,\mu$m; see below for a definition and see
appendix D for more details.  Flux densities in columns 3, 4, 6 and 7
have been corrected for the point spread function being larger than
the pixel size of the detector and for Rayleigh-Jeans
colour-correction (cc); for ISO fluxes, the inband flux has been
divided by 0.69 (the correction for the point spread function (=psf),
see Appendix B) and by 1.06 (cc) and for IRAS fluxes, the IFSC or IPSC
flux have been divided by 1.31 (cc). The ``disk emission'' in column 8
is ``de-colour-corrected'' from column 6.

We have checked the quality of our results at 60\,\um\  in two ways:
(i) by comparing ISO with IRAS flux densities; (ii) by comparing
fluxes measured by ISO with predictions based on the $(B-V)$
photometric index. The second approach allows us to assess the quality
of ISO flux densities below the IRAS sensitivity limit.

\begin{figure}
\psfig{figure=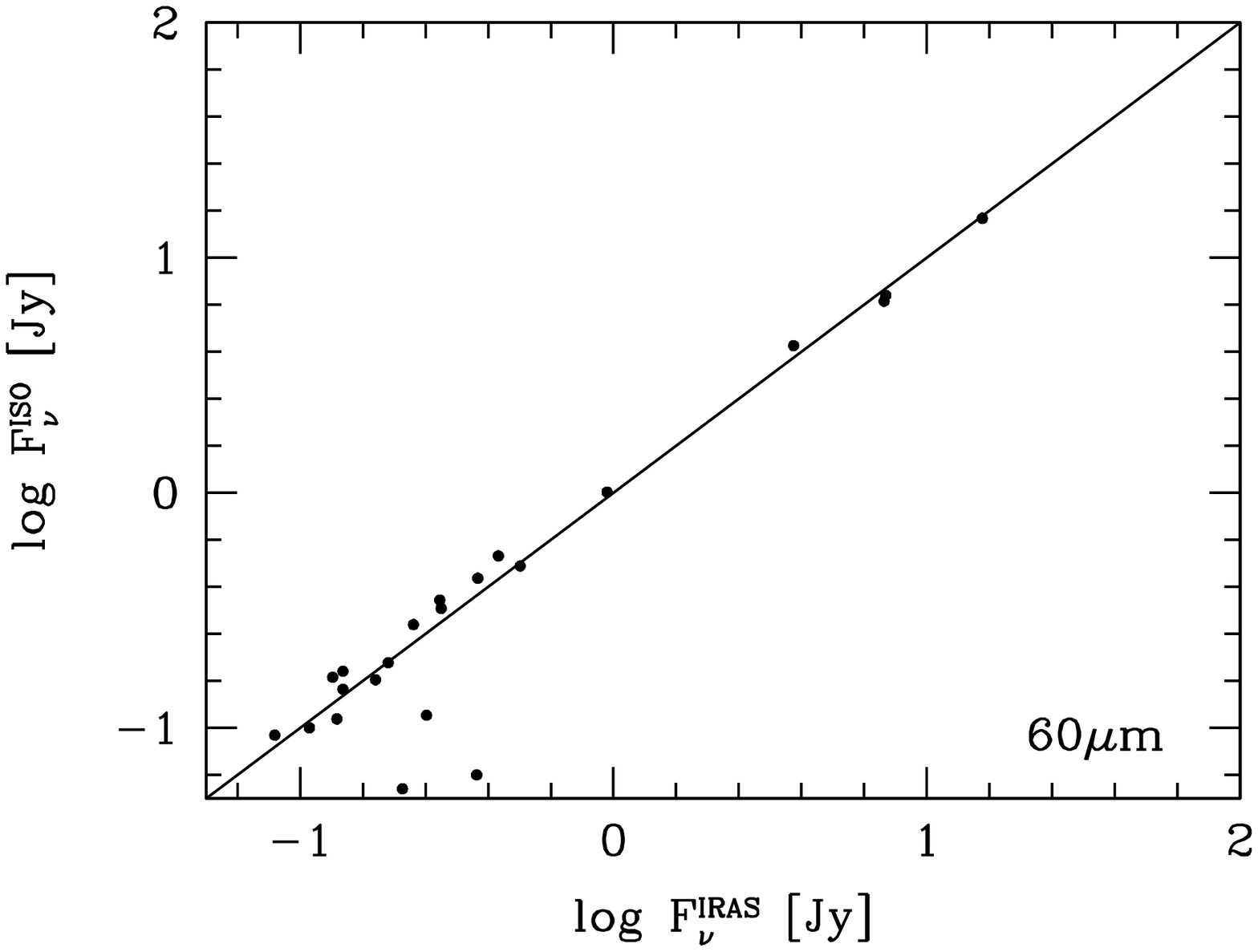,width=8.8cm,clip=}
\caption{\label{fig:isoiras} Correlation of fluxes measured by IRAS
and by ISO, respectively. The line marks the relation
  $F_{\nu}^{\mathrm{IRAS}}=F_{\nu}^{\mathrm{ISO}}$.}
\end{figure}
\begin{figure}
\psfig{figure=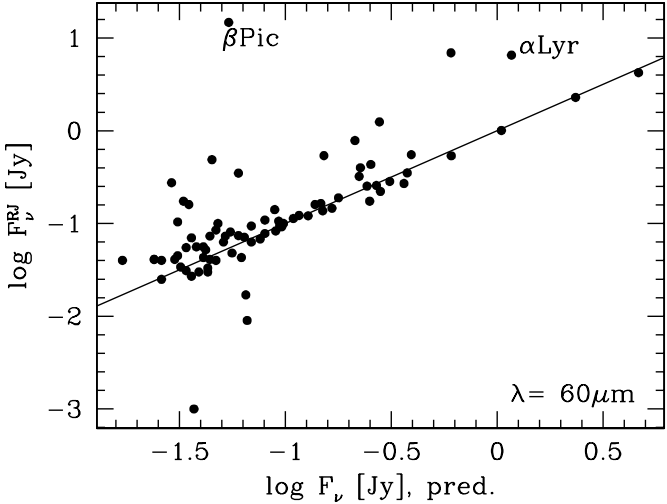,width=8.8cm,clip=}
\caption{\label{fig:measpred60}  Diagram of predicted and measured 
  fluxes at 60\,\um.  The predicted fluxes were derived mainly from
  Eq. (\ref{eq:plets}) except in a few cases where Kurucz model
  atmospherese were fitted to photometric points at optical
  wavelengths.  See text. The line marks the relation $F_{\nu}=
  F_{\nu}^{\mathrm{pred}}$.}
\end{figure}
\begin{figure}
  \psfig{figure=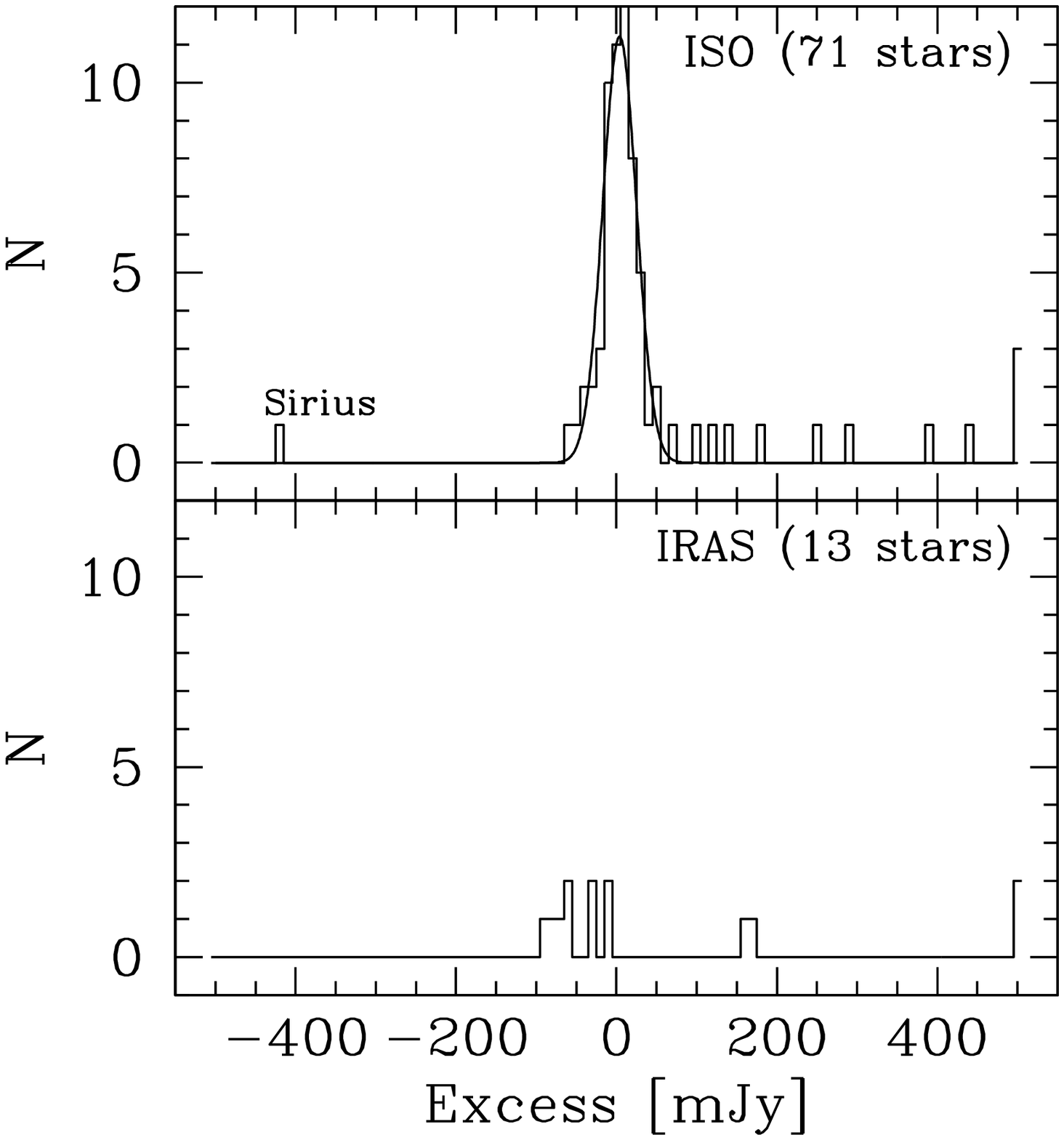,width=8.8cm,clip=,angle=270}
\caption{\label{fig:histex60} Histogram of the differences between the 
  measured flux density and the one predicted at 60\,\um. Top:
  distribution of the flux densities measured by ISO; there are three
  stars with an excess higher than 500\,{mJy}; the drawn curve is a
  Gauss curve with average $\mu =\, 4$mJy and dispersion $\sigma =\,
  21$mJy.  Bottom: the same for stars where only IRAS data are
  available; two stars have an excess higher than 500\,{mJy}.}
\end{figure}

\subsection{The correlation between IRAS and ISO measurements at 60$\,\mu$m}

\begin{figure}
\psfig{figure=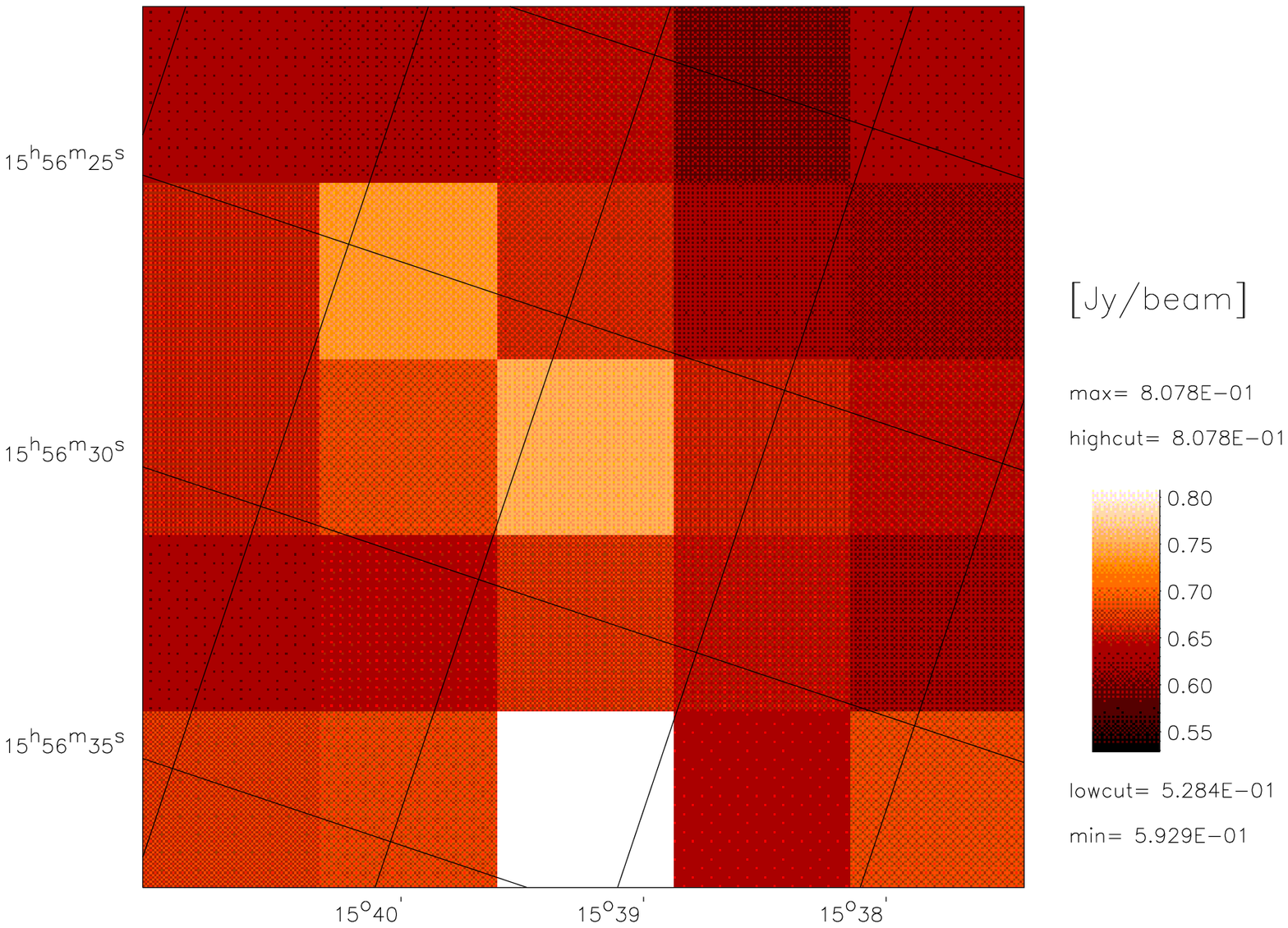,height=5.0truecm,width=5.7truecm}
  \caption{The 60\,$\mu$m image in spacecraft orientation of the
  region around HD 142860 as obtained from the ISOPHOT minimap.
  There are three point sources in the field, the position of the
  source in the centre corresponds to the position of HD142860. The
  upper source has $F_{\nu}=$ 140$\pm$40\, mJy and coordinates (J2000)
  RA ${\rm 15^h 56^m 25^s}$, Dec $15^\circ 40' 43''$; the lower
  source has $F_{\nu}=$ 250$\pm$40\, mJy and coordinates
  (J2000) RA ${\rm15^h 56^m 33^s}$, Dec. $15^\circ 39' 15''$.}
  \label{fig:HD142860}
\end{figure}

Fig. \ref{fig:isoiras} shows the strong correlation between IRAS and
ISO 60\,\um\  flux densities down to about the 60\, mJy level of ISO.  For
three of the fainter stars the IRAS fluxes are considerably higher
than those of ISO. For one of these three, HD 142860, the ISO
measurements show the presence of two nearby 60 $\mu$m sources; the
larger IRAS beam has merged the three sources; see Fig.
\ref{fig:HD142860}. For the two remaining sources in Fig.
\ref{fig:isoiras} with different IRAS and ISO flux densities we assume
that the IRAS measurement is too high because noise lifted the
measured flux density above the detection limit, a well-known effect
for measurements close to the sensitivity limit of a telescope.

\subsection{The correlation between predicted and measured flux densities
at 60\,$\mu$m}

A strong correlation exists between predicted and measured flux
density, columns (8) and (3) in Table \ref{tab:60um}, as seen in
Fig. \ref{fig:measpred60}.

Fig. \ref{fig:histex60} shows that the distribution of the excess
fluxes can be split into two components: a very narrow distribution
around zero plus a strong wing of positive excesses, i.e.  cases where
we measure more flux than is produced by the stellar photosphere. In
these cases a disk is very probably present. A Gauss curve has been
drawn in the figure with parameters $\mu=4$\, mJy and $\sigma=21$\,
mJy, where $\mu$ is the average and $\sigma$ the dispersion. The value
of $\sigma$ agrees with the magnitude of individual error measurements
as given in column 8 in Table \ref{tab:60um}. The one case in Fig.
\ref{fig:histex60} with a strong negative excess, i.e. where we
measured less than is predicted, concerns Sirius, $\alpha$ CMa; we
interpret this negative excess as a consequence of the poor correction
for transient effects of the detector for this strong infrared source.

We have concluded that a disk is present when $F_{\nu}^{\mathrm{exc}}
>\mu+3\sigma= 65$ mJy. A summary of data on all stars with disks is
in Table \ref{tab:summary}. HD\,128167 has also been labelled as a
``detection'' although $F_{\nu}^{\mathrm{exc}}$ is only
$2.8\,\sigma_\nu$.  The star is one of the few with a detection at
170\,\um\  and this removed our doubts about the detection at 60\,\um.

\subsection{Results at 170$\mu$m}

The results are shown in Table \ref{tab:170um} which has the same
structure as Table \ref{tab:60um}. IRAS did not measure beyond 
100\,\um\ and we have no existing data to compare with our ISO data.
Neither can we make a useful comparison between measured and predicted
flux densities because the photospheric flux is expected to be roughly
1/8 of the 60\,\um\  flux density and for most stars this is below the
sensitivity limit of ISO.

We accepted fluxes as real when $F^{\mathrm{exc}}_{\nu}>
3\sigma_{\nu}$ \textit{and} the minimap showed flux only in the pixel
illuminated by the star.  This leads to seven detections at 170\,\um\ 
in Table \ref{tab:170um}. All seven stars have excess emission also at
60\,\um\  except HD 20807 and HD 38393. Very probably these last two
stars have accidentally been misidentified with unrelated background
sources, as we will show now.

\subsection{Should some detections be identified with unrelated 
field sources?}

We need to consider the possible influence of the field-source
population upon our results.

\citet{dole:00}, \citet{mats:00}, \citet{oliv:00} and \citet{elba:00}
quote source counts at respectively 170, 90 and 15\,\um\  from which
the surface density of sources on the sky can be read down to the
sensitivity limits of our measurements. In the 170 and 90\,\um\   cases
this involves the authors' extrapolations, via models, of their source
counts from the roughly 100-200\, mJy flux limits of their respective
datasets. At 60\,\um\  source counts can be approximated, with
sufficient accuracy for present purposes, by interpolation from the
other wavelengths. Using these source densities we now estimate
the probability that our samples of detections contain one or two
field sources unrelated to the star in question.

Since we know, for all of our targets, into exactly which pixel of the
PHT map they should fall, we need to consider the probability that a
field source with flux down to our sensitivity limit falls into the
relevant PHT pixel. This effective ``beam'' area is $45''\times 45''$ and
$100''\times 100''$, at 60 and 170$\,\mu$m respectively.

At 60\,\um\  we have explored 84 beams (targets) and at 170\,\um\  52
beams (targets).  We apply the binomial distribution to determine the
probability $P(q,r)$ that at least $r$ spurious detections occur in
$q$ trials when the probability per observation equals $p$.

Table \ref{tab:chance} lists the following parameters: column(1) shows
the wavelength that we consider; column (3) contains the probability
$p$ to find a source in any randomly chosen pixel with a flux density
above the limit $F_{\nu}^{\mathrm{lim}}$ given in column (2). Column
(4) lists the number $q$ of targets (i.e. trials) in the 60 and 170\um
\,  samples. Columns (5)-(7) give the probability $P(q,r)$ of
finding at least 1, 2 and 3 spurious detections with $F_{\nu}
>F_{\nu}^{\mathrm{lim}}$ within a sample of size $q$.

\begin{table}
\caption{\label{tab:chance} Probability of spurious detections}
\footnotesize
\begin{tabular}{|rrrrrrr|}
\hline
&&&&&&\\
$\lambda$&$F_{\nu}^{\mathrm{lim}}$ &$p$& $q$&$P(q,1)$&$P(q,2)$&$P(q,3)$\\
\um & mJy & & & & & \\
\hline
  60&    100&   0.006& 84 & 0.397&  0.091&\\
  60&    150&  0.0015& 84 & 0.117 &  0.007& \\
  60&    200&  0.001 & 84 &  0.079 &  0.003& \\
 170&     50&  0.15  & 52 &  1.000 &  0.997 &  0.988 \\
 170&    100&  0.07  & 52 &  0.975 &  0.881 &  0.702 \\
 170&    200&  0.04  & 52 &  0.875 &  0.61 &  0.334 \\
 170&    300&  0.005 & 52 &  0.226&  0.027 &\\
 170&   1000&  0.00025 &52 &  0.013&& \\

\hline
\end{tabular}
\end{table}

\subsubsection{Probability of spurious detections at 60\,\um}

It follows from Table \ref{tab:chance} that there is a 40\% chance
that at least one of the two detections at 60$\,\mu$m below 100 mJy is
due to a field source, and there is a 9\% chance that both are.

If we ignore field sources and consider the likelihood of spurious
excesses occurring above a $3\,\sigma$ detection-limit due purely to
statistical fluctuations in the measurements, we find that random noise
contributes (coincidentally) a further 0.006 spurious detections per
beam, on average, for the faintest detections (near $3\,\sigma$).

The cumulative probability, therefore, is 0.64 that at least 1 of the two
detections at 60$\,\mu$m below 100 mJy is not related to a disk; the
probability is 0.17 that they are both spurious.

\subsubsection{Probability of spurious detections at 170\,\um}

Table \ref{tab:170um} lists 3 detections below 100 mJy. Two (HD 20630
and 38393) have not been detected at 60\,\um.  Table \ref{tab:chance}
shows that the probability is high that both are background sources
unrelated to the two stars in question. In the further discussion
these two stars have been considered to be without a disk. The third
source with a 170\,\um\  flux density below 100 mJy, HD 128167, has been
detected also at 60\,\um. We assume that this detection is genuine
and that the source coincides with the star. The remaining four
detections at 170\,\um\  with $F_{\nu}> 100$ mJy are also correctly
identified with the appropriate star.

\begin{table*}
\caption{\label{tab:summary} The 60$\mu$m excess stars}
\scriptsize
\begin{center}
\begin{tabular}{|rrrr|rrrr|rrr|r|}
\hline
    &      &        &      &
\multicolumn{4}{c|}{60\,$\mu$m} &
\multicolumn{3}{c|}{170\,$\mu$m}& \\
 HD & Name & Spect. & age  &
$F_{\nu}$  &
$\sigma_{\nu}$  &
$F_{\nu}^{\mathrm{disk}}$ &
$^{10}$log$\tau_{60}^{\mathrm{disk}}$          &
$F_{\nu}$  &
$\sigma_{\nu}$  &
$F_{\nu}^{\mathrm{disk}}$ &
$M_{\rm d}$ \\
%$T_{60/170}$\\
    &     &     & Gyrs & mJy & mJy & mJy & & mJy & mJy & mJy & $10^{-5}M_{\oplus}$ \\
%   &     &     &      & cc(RJ) & cc(RJ) &  & cc(RJ) & cc(RJ) &  \\
(1)&(2)&(3)&(4)&(5)&(6)&(7)&(8)&(9)&(10)&(11)&(12)\\
\hline
10700 &  $\tau$ Cet    & G8V     & $7.24$ & $433$ & $37$ & $190$      & $-4.6$  & $125$ & $21$ & $120$ & $20$ \\
17925 &                & K1V     & $0.08$ & $104$ & $24$ & $80$       & $-3.9$ & &  &  &  $108$  \\
22049 & $\epsilon$ Eri & K2V     & $0.33$ & $1250$ & $100$ & $1260$   & & & & & \\
30495 & $58$ Eri       & G3V     & $0.21$ & $174$ & $31$ & $150$      & $-4.1$ & $51$ & $25$ &  & $73$   \\
38678 & $\zeta$ Lep    & A2Vann  & $0.37$ & $349$ & $22$ & $310$      & $-4.7$ & $22$ & $48$ &  &  $18$ \\
39060 & $\beta$ Pic    & A3V     & $0.28$ & $14700$ & $346$ & $15500$ & $-2.8$ & $3810$ & $143$ & $4600$ & $1200$\\
75732 & $\rho^1$ Cnc   & G8V     & $5.01$ & $160$ & $28$ & $130$      & $-3.8$ &  &  &  & $130$ \\
95418 & $\beta$ UMa    & A1V     & $0.36$ & $539$ & $135$ & $410$     & $-5.0$ & $133$ & $290$ &  & $8$  \\
102647 & $\beta$ Leo   & A3V     & $0.24$ & $784$ & $100$ & $750$     & $-4.8$ &  &  &  & $13$ \\
128167 & $\sigma$ Boo  & F3Vwvar & $1.70$ & $100$ & $19$ & $55$       & $-5.0$ & $56$ & $12$ & $60$ & $8$ \\
139664 & g Lup         & F5IV-V  & $1.12$ & $488$ & $48$ & $470$      & $-4.0$ & $122$ & $830$ &  & $84$  \\
172167 & $\alpha$ Lyr  & A0Vvar  & $0.35$ & $6530$ & $217$ & $5700$   & $-4.8$ & $2620$ & $142$ & $3000$ & $13$ \\
207129 &               & G2V     & $6.03$ & $275$ & $55$ & $260$      & $-3.8$ & $293$ & $23$ & $350$ & $132$ \\
216956 & $\alpha$ PsA  & A3V     & $0.22$ & $6930$ & $204$ & $6700$   & $-4.3$ &  &  &  & $44$ \\
\hline
\end{tabular}
\end{center}
\footnotesize
{Note: Fluxes in columns 5,6,9,10 are corrected for point spread function and Rayleigh-Jeans colour-correction.
Excesses in columns 7 and 11 are ``de-colour-corrected'' from columns 5 and 9
respectively (see text).}
\normalsize
\end{table*}

\section{Discussion}

Data for the stars with a disk are summarized in Table
\ref{tab:summary}. Each column contains quantities defined and used in
earlier tables with the exception of the last column that contains
an estimate of the mass of the disk, \Mdisk.

\subsection{Comparison with the IRAS heritage}

The overall good agreement between the IRAS and ISO measurements has
already been discussed. The IRAS data base has been explored by
several different groups in search of more Vega-like stars.
\citet{back:93} have reviewed most of these searches. Their table X
contains all stars with disks. For nine of those in table X we have
ISO measurements, and in eight cases these show the presence of a
disk. The one exception is $\delta$ Vel, HD 74956. We find excess
emission but the excess is insignificant (only $1.7\sigma$) and we did
not include the star in table \ref{tab:summary}. Thus our data agree
well with those discussed by \citet{back:93}. Although ISO was more
sensitive than IRAS at 60\,\um\ by about a factor of 5 we have only
one new detection of a disk: HD 17925, with an excess of 82 mJy at 60
\um .

\citet{plet:99} have analysed IRAS data in search of ``Vega-like''
stars and paid special attention to our list of candidate stars. In
general there is good agreement between their conclusions and ours.
Nine of the disk stars they find in IRAS we confirm with our ISO data.
Four of the stars for which they find evidence of a disk at 60\,\um\ 
are without excess emission in our sample. In two cases (HD 142860 and
HD 215789) the difference IRAS/ISO is large: IRAS: 413 and 220 mJy;
ISO: 113 and 78 mJy). In both cases we are of the opinion that at such
low flux levels the ISO data are to be preferred over the IRAS data.

\subsection{Differences with disks around pre-main-sequence stars}
Disks have been found around many pre-main-sequence (=PMS) stars; here
we discuss disks around main-sequence (=MS) stars. Disks around PMS
stars are always detected by their molecular line emission; they
contain dust and gas. The search for molecular emission lines in MS
disks, however, has been fruitless so far \citep{lise:99}. This is in
line with model calculations by \citet{kamp:00} who show that CO in
disks around MS-stars will be dissociated by the interstellar
radiation field. Recent observations with ISO indicate the presence of
H$_2$ in the disk around $\beta$ Pic and HST spectra show the presence
of CO absorption lines (van Dishoeck, private communication), but the
disk around $\beta$ Pic is probably much ``fatter'' than those around
our MS-stars; it is not even certain that $\beta$ Pic is a PMS or
MS-star.  We will henceforth assume that disks detected around
MS-stars contain only dust and no gas.

\subsection{A simple quantitative model}

We have very little information on the disks: in most cases only the
photometric flux at 60\,\um . For the quantitative discussion of our
measurements we will therefore use a very simple model. We assume a
main-sequence star with an effective temperature \Teff\ and a
luminosity \Lstar. The star is surrounded by a disk of $N$ dust
particles. For simplicity, and to allow an easy comparison between
different stars, we use a unique distance of the circumstellar dust of
$r=50$ AU.  This value is consistent with the measurements of
spatially resolved disks like Vega and $\epsilon$\, Eri and also with
the size of the Kuiper Belt in our own solar system.  The particles
are spherical, have all the same diameter and are made of the same
material. The important parameter of the disk that varies from star to
star is $N$. The temperature, \Tdust, of each dust particle is
determined by the equilibrium between absorption of stellar photons
and by emission of infrared photons; thus \Tdust\ depends on \Teff .

Each dust particle absorbs photons with an effective cross section
equal to $Q_\nu \pi a^2$; $Q_\nu$ is the absorption efficiency of the
dust and $a$ the radius of a dust particle. The average of $Q_\nu$
over the Planck function will be written as $Q_{\mathrm{ave}}$.  The
dust particles absorb a fraction $\tau\equiv N Q_{\mathrm{ave}}
(T_{\mathrm{eff}})\, a^2/(4\, r^2)$ of the stellar energy and reemit
this amount of energy in the infrared; in all cases the value of
$\tau$ is very small.  We will call $\tau$ the ``optical depth of the
disk''; it represents the extinction by the disk at visual wavelengths:
 
\begin{equation}
\tau=\frac{L_{\mathrm{d}}}{L_{*}}= \frac{F^{\mathrm{d}}_
{\mathrm{bol}}}{F^{\mathrm{pred}}_{\mathrm{bol}}}.        
\end{equation} 

The mass of the disk, $M_{\mathrm{d}}$, is proportional to $\tau$:
\begin{equation}
M_{\mathrm{d}}= \frac{16\pi}{3}\,\frac{\rho\, a\, r^2}{Q_{\mathrm{ave}}}
\tau
\end{equation}
We will use spheres with a radius $a$ of 1\,\um\  and with material
density $\rho$ and optical constants of interstellar silicate
\citep{drai:84}; we use $Q_{\mathrm{ave}}=\, 0.8$ and derive \Mdisk$=
0.5\,\tau\,$M$_\oplus$.  It is known that the grains in Vega-like
systems are much larger than interstellar grains.  For A stars, the
emission is probably dominated by grains larger than 10\,\um\ 
\citep{auma:84, zuck:93, chin:91}, because smaller grains are blown
out by radiation pressure.  However, for F, G, and K stars, the
blowout sizes are 1\,\um\ or smaller and it must be assumed that the
emission from these stars is dominated by smaller grains.  We
calculate mass estimates using the grain size of 1\um.  Since the mass
estimates depend linearly upon the grain size, the true masses of
systems with bigger grains can easily be calculated by scaling the
value.  The mean absorption efficiency factor $Q_{\mathrm{ave}}$ is
only weakly dependent upon the grain size for sizes between 1 and
100\um.

\begin{figure}
\psfig{figure=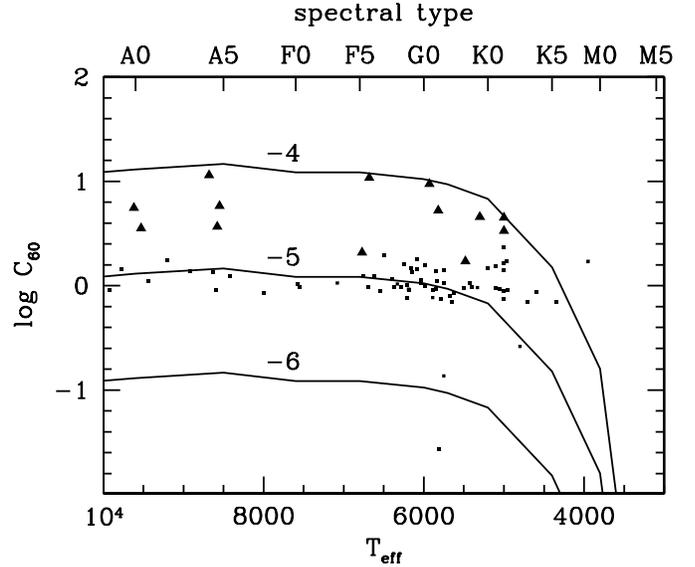,width=8.8cm,clip=}
\caption{\label{fig:Teff-C60}  The $^{10}$logarithm of the fraction of 
  stellar energy emitted by the disk is shown as a function of \Teff
  and $C_{60}$. The labels to the curves indicate the $^{10}$logarithm
  of $\tau$. The triangles represent stars with a disk, and the small
  squares indicate the upper limit of non-detections. Using our
  standard dust particle model the mass of each disk is given by
  \Mdisk$= 0.5\, \tau\, $M$_\oplus$ (see text).}
\end{figure}

Numerical simulations made us discover a simple property of this model
that is significant because it makes the detection probability
constant for disks of stars of different spectral type . Define a
variable called ``contrast'': $ C_{60}\equiv (L_{\nu,\mathrm{d}}
/L_{\nu,*})_{60\, \mu m}$, and assume black body radiation by the star
and by the dust particles, then $C_{60}$ is constant for \Teff\, in
the range of A, F, and G-stars. The reason for this constancy is that
when \Teff\, drops the grains get colder and emit less in total, but
because 60\,\um\ is at the Wien side of the Planck curve, their
emission rate at 60\,\um\ goes up. For a more elaborate discussion see
appendix C.  Let us then make a two-dimensional diagram of the values
of $\tau$ (or of $M_{\mathrm{d}}$) as a function of \Teff\ and
$C_{60}$: see Fig.  \ref{fig:Teff-C60}.  Constant values of $\tau$
appear as horizontal contours for \Teff$\, > 5,000$ K. The triangles
in the diagram represent the disks that we detected; small squares
represent upper limits. The distribution of detections and upper
limits makes clear that \textit{we detected all disks with
  $\tau>2\times 10^{-5}$ or $M_{\mathrm{d}}\,1.0\times
  10^{-5}$M$_\oplus$ around the A, F, G-type stars in our sample of 84
  stars}; we may, however, have missed a few disks around our K stars
and we may have missed truncated and thus hot disks.

\subsection{The incidence and survival of remnant dust disks}

The results discussed here have also been presented in
\citet{habi:99}.

Stellar ages have been derived in an accompanying paper
\citep{lach:99}. Errors in the determination of the ages have been
given in that paper; occasionally they may be as large as a factor of
2 to 3; errors that large will not detract from our main conclusions.

\subsubsection{The question of completeness and statistical bias}

Our sample has been selected from the catalogue of stars within 25\,
pc from the Sun by \citet{wool:70}; this catalogue is definitely
incomplete and so must be our sample. Even within the distance limits
given in Table \ref{tab:distlimits} stars will exist that we could
have included but did not. This incompleteness does not, however,
introduce a statistical bias: we have checked that for a given
spectral type the distribution of the stellar distances is the same
for stars with a disk as for stars without a disk; this is illustrated
by the average distances in Table \ref{tab:distances}.

\begin{table}[h!]
\caption{\label{tab:distances}: Average distances of stars with and
without a disk} 
\small
\begin{tabular}{|c|cc|cc|}
\hline
&{\#}&\multicolumn{1}{c|}{without disk}&{\#}&\multicolumn{1}{c|}{with disk}\\
&&\multicolumn{1}{c|}{(pc)}&&\multicolumn{1}{c|}{(pc)}\\
\hline
A**&9& 22.6$\pm$11.0 &6& 16.8$\pm$7.1\\
F**&21& 14.6$\pm$4.6 &2& 14.0$\pm$...\\
G**&17& 12.3$\pm$3.9 &4& 11.3$\pm$5.3\\
K**&20& 9.5$\pm$3.8 &2& 6.6$\pm$... \\
\hline
\end{tabular}
\end{table}

\subsubsection{Detection statistics and stellar age}

\begin{figure}
\psfig{figure=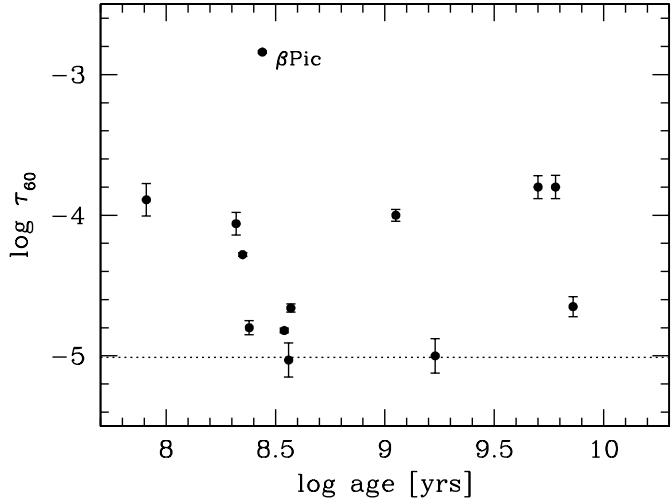,width=8.8cm,clip=}
\caption{\label{fig:tau-of-age}  $\tau$, the fraction of the stellar light 
  reemitted at infrared wavelengths, is shown as a function of stellar
  age}
\end{figure}

\begin{table}[h!]
\caption{\label{tab:detstat}: Detection statistics}
\small
\begin{tabular}{|c|cc|cc|cc|cc|}
\hline
&\multicolumn{2}{|c|}{$<400$}&\multicolumn{2}{c|}{400-1000}
&\multicolumn{2}{c|}{1.0-5.0}&\multicolumn{2}{c|}{$>5.00$}\\
&\multicolumn{2}{|c|}{Myr}&\multicolumn{2}{c|}{Myr}
&\multicolumn{2}{c|}{Gyr}&\multicolumn{2}{c|}{Gyr}\\
& tot & disk & tot & disk & tot & disk & tot & disk \\
\hline
A**& 10 & 6 & 4 & 0 & 1 & 0 & 0 & 0 \\ 
F**& 0 & 0 & 1 & 0 & 17 & 2 & 5 & 0 \\ 
G**& 2 & 1 & 0 & 0 & 7 & 0 & 12 & 3\\
K**& 3 & 2 & 2 & 0 & 5 & 0 & 12 & 0\\
& & & & & & & & \\
total& 15 & 9 & 7 & 0 & 30 & 2 & 29 & 3\\
\hline
\end{tabular}
\end{table}

Fig. \ref{fig:tau-of-age} presents graphically the fraction of the
(visual) stellar light reemitted in the infrared by the disk as a
function of the stellar age. Similar diagrams based mainly on IRAS
results, have been published before- see, for example,
\citet{holl:98}. A general, continuous correlation appears: disks
around PMS-stars (e.g. Herbig AeBe) are more massive than disks around
stars like $\beta$ Pic and Vega, and the disk around the Sun is still
less massive. These earlier diagrams have almost no data on the age
range shown in Fig. \ref{fig:tau-of-age} and the new ISO data fill in an
important hole.

Table \ref{tab:detstat} summarizes the detections at 60 \um\ 
separately for stars of different age and of different spectral type
together with the same numbers for stars with a disk; in the column
marked ``tot'' the total number of stars (disks plus no-disks) is
shown and under the heading ``disk'' the number of stars with a disk.
The total count is 81 instead of 84 because for three of our target
stars (two A-stars and one K-star) the age could not be estimated in a
satisfying manner. Table \ref{tab:detstat} shows that the stars with a
detected disk are systematically younger than the stars without disk:
out of the 15 stars younger than 400 Myr nine (60\%) have a disk; out
of the 66 older stars only five have a disk (8\%).  Furthermore, there
exists a more or less sharply defined age above which a star has no
longer a disk. This is best demonstrated by the A-stars. Six A-stars
have a disk; the stellar ages are 220, 240, 280, 350, 360, 380 Myr.
For the A-stars without disk the corresponding ages are 300, 320, 350,
380, 420, 480, 540, 890, 1230 Myr: 350 to 400 Myr is a well-defined
transition region. We conclude that the A stars in general arrive on
the main-sequence with a disk, but that they loose the disk within 50
Myr when they are about 350 Myr old.

Is what is true for the A-stars also valid for the stars of other
spectral types? Our answer is ``probably yes'': of the five F, G, and
K stars younger than 400 Myr three (60\%) have a disk. Of the 61 F, G,
and K stars older than 400 Myr five have a disk (one in twelve or
8\%). The percentages are the same as for the A-stars but the 60\% for
young G- and K-stars is based on only three detections. It seems that
the disks around F, G, and K stars decay in a similarly short time
after arrival on the main sequence.

\begin{figure}
\psfig{figure=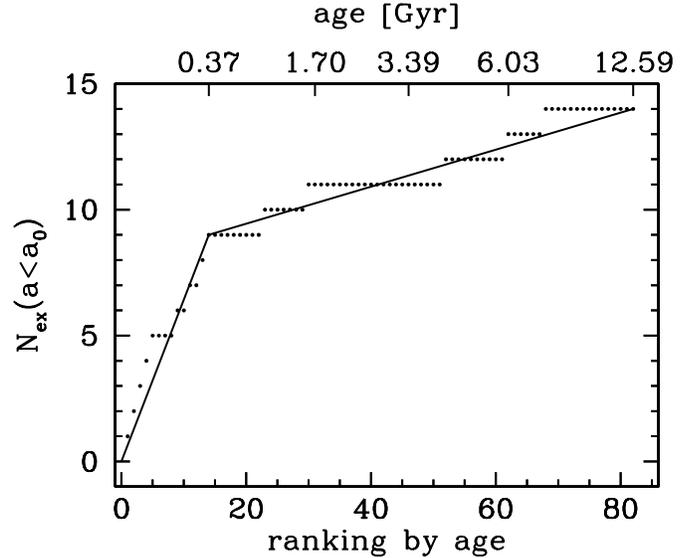,width=8.8cm,clip=,angle=270}
\caption{\label{fig:cumul_just_model} Cumulative distribution of excess
  stars, as a function of the index after sorting by age. The two
  segments of a continuous straight line are predicted by assuming
  that in the first 400 Myr the rate of disappearance of disks is much
  higher than afterwards (see text).}
\end{figure}

An immediate question is: do \textit{all} stars arrive at the
main sequence with a disk? Studies of pre-main-sequence stars show
that disks are common, but whether they always exist is unknown. The
sequence of ages of the A-stars shows that the three youngest A-stars
have a disk. This suggests that all stars arrive on the main sequence
with a disk, but the suggestion is based on small-number statistics.
We therefore leave the question without an answer but add two relevant
remarks without further comment: some very young stars have no
detectable disk, for example HD 116842 (A5V, 320 Myr), HD 20630 (G5V,
300 Myr), HD 37394 (K1V, 320 Myr) and some old stars have retained
their disk; examples: HD 10700 (G8V, 7.2 Gyr), HD 75732 (G5V, 5.0 Gyr)
and HD 207129 (G0V, 4.4 Gyr); the last case has been studied in detail
\citep{jour:99}. 

The age effect is shown graphically in
Fig.~\ref{fig:cumul_just_model}; it displays the cumulative
distribution of stars with a disk. The x-axis is the index of a star
after all stars have been sorted by age.  At a given age the local
slope of the curve in this diagram gives the probability that stars of
that age have a disk. The two line segments shows how the cumulative
number increases when 70\% of the disks disappear gradually in the
first 0.4 Gyr and the remaining 30\% gradually in the 12 Gyr thereafter. 

In section 5.6 we will review the evidence that at about 400 Myr after
the formation of the Sun a related phenomenon took place in the solar
system.

\subsection{The need to continuously replenish the dust particles}

In ``Vega-like'' circumstellar disks the dust particles have a
life-time much shorter than the age of the star. Within 1 Myr they
will disappear via radiation pressure and the Poynting-Robertson
effect \citep{auma:84}. An upper limit of $10^6$ year is given for
dust around A-type stars by Poynting-Robertson drag
\citep{burn:79,back:93}; the actual life time will be smaller: for
$\beta$ Pic, \citet{arty:97} find only 4000 years. Continuously new
grains have to replace those that disappear. A plausible mechanism
that can supply these new grains at a high enough rate and for a
sufficient long time are the collisions between asteroids and
planetesimals.  Direct detection of such larger bodies is not yet
possible although the existence of comets around $\beta$ Pic is
suggested by the rapidly appearing and disappearing components of the
CaII K-absorption line \citep{ferl:87}. The total mass of the dust
that ISO detected is about that of the Moon. To produce the dust for
400 Myr much more mass must be present in invisible form, that of
asteroids or ``planetesimals''.  Thus the disappearance of the
infrared excess on a timescale of 400\, Myr does not trace the removal
of the dust grains, but \emph{the lifetime of the disk or
  planetesimals that replenishes the dust}. In the solar system the
same may have happened; see below.

\subsection{Disks in the presence of a companion star or a planet}

When a star has a companion or a planet the gravitational field will
have a time-variable component. Will this component destroy the disk?
Not necessarily so: the planets Jupiter and Saturn have both a dust
disk and many satellites.

On purpose we did not select narrow binaries: we rejected stars within
1 arcmin from a target star, unless this other star was at least 5
magnitudes fainter in the $V$-band. This criterium accepts wide
multiple-stars and indeed these occur.  We used the Hipparcos
Catalogue to check all 84 stars from Table \ref{tab:stars} for
multiplicity. Forty-eight stars have an entry in the ``Catalogue of
companions of double and multiple stars'' (=CCDM), see
\citet{domm:94}.  Among the 14 stars with a disk there are seven wide
multiple-stars. In one case (HD22049) the star is part of an
astrometric double; we ignore the object.  That leaves us with six
stars that have both a disk and stellar companions. The conclusion is
therefore that companions do not necessarily destroy a disk.

Table \ref{tab:multiple} contains information on these six stars with
both a disk and (at least) one companion. In column (1) the name
appears, in column (2) the HD-number and in column (3) the
entry-number in the CCDM; column (4) gives the total number of
companions given in the CCDM, column (5) gives the distance, $r$, between A
and B in astronomical units and column (6) the magnitude difference in
the $V$-band between the first and the second component (``A'' and
``B'', respectively).

There are at least two remarkable cases in Table \ref{tab:multiple}.
One is Vega (HD 172167) with four companions; its brightest companion is
at 490 AU, but its closest companion at only 200 AU, just outside of
Vega's disk. The other is $\rho^1$ Cnc that has a disk \citep{domi:98,
  tril:98, jaya:00}, a planet \citep{butl:97} and a stellar
companion.

\begin{table}[h!]
\caption{\label{tab:multiple} Multiple systems with a dust disk}
\small
\begin{tabular}{|rrr|rrr|}
\hline
Name&HD & CCDM & $N_{\mathrm{tot}}$ & $r$ & $\Delta m$\\
&&&&AU&\\
\hline
$\tau$ Cet   & 10700  & 01441-1557     & 1 &  328 &  9.5\\
$\rho^1$ Cnc & 75732  & 08526+2820     & 1 & 1100 &  7.2\\
$\beta$ Leo  &102647  & 11490+1433     & 3 &  440 & 13.5\\
$\sigma$ Boo &128167  & 14347+2945     & 2 & 3700 &  5.3\\
$\alpha$ Lyr &172167  & 18369+3847     & 4 &  490 &  9.4\\
             &207129  & 21483-4718     & 1 &  860 &  3.0\\
\hline
\end{tabular}
\end{table}

The data in Table \ref{tab:multiple} thus
show that disks are found in wide multiple-systems: multiplicity does
not necessarily destroy a disk.

\subsection{The connection to the solar system}

The solar system shows evidence for a fast removal of a disk of
planetesimals a few hundred Myr after the Sun formed a disk. The best
case is given by the surface of the Moon, where accurate crater
counting from high resolution imaging can be combined with accurate
age determinations of different parts of the Moon's surface.  The age
of the lunar surface is known from the rocks brought back to earth by
the Apollo missions; the early history of the Moon was marked by a
much higher cratering rate than observed today; see for a discussion
\citet{shoe:99}. This so-called ``heavy bombardment'' lasted until some
600\, Myr after the formation of the Sun. Thereafter the impact rate
decreased exponentially with time constants between 10$^7$ and a few
times 10$^8$ years \citep{chyb:90}. 

Other planets and satellites with little erosion on their surface
confirm this evidence: Mercury \citep{stro:88}, Mars \citep{ashx:96,
  sode:74} Ganymede and Callisto \citep{shoe:82,neuk:97,zahn:98}. The
exact timescales are a matter of debate. Thus there are indications of
a cleanup phase of a few hundred Myr throughout the solar system;
these cleanup processes may be dynamically connected.

\section{Conclusions on the incidence of remnant disks}

The photometers on ISO have been used to measure the 60 and 170
$\mu m$ flux densities of a sample of 84 main-sequence stars with
spectral types from A to K.

On the basis of the evidence presented we draw the following conclusions: 
\begin{itemize}
\item Fourteen stars have a flux in excess of the expected
  photospheric flux. We conclude that each of these has a
  circumstellar disk that is a remnant from its pre-main-sequence
  time. Two more stars may have a disk, but there is a significant
  chance that the emission is due to a background galaxy.
\item The overall incidence of disks is 14/84 or 17\%. A-stars have a
  higher incidence than the other stars.
\item We prove that the detectability of a given disk is the same for
  A, F and G-stars with the same photospheric flux at 60\um ; K stars
  have a lower probability of detection.
\item The disks that we detect have a value of $\tau$, between
  $2\times 10^{-4}$ and $4\times 10^{-6}$. The upper limit is real:
  main-sequence stars do not carry stronger disks; the exception is
  $\beta$ Pic with $\tau=8\times 10^{-3}$. The lower limit is caused
  by selection effects: fainter disks are below our detection
  threshold.
\item Six out of the ten A-type stars younger than 400 Myr have a
  disk; the disk is absent around all five older A-type stars: the
  disks disappear around this age.
\item Disks around F-, G- and K-stars probably disappear on a similar
  time scale. The disappearance of disks is \textit{not} a continuous
  process; 400 Myr is the age at which most disks disappear.
\item In the history of our solar system the abrupt ending of the
  initial ``heavy bombardment'' of the Moon has the same time scale.
\item We suggest that the disks that we detect are actually sites
  where a ``heavy bombardment'' takes place now. The time scale on
  which the disks disappear is actually the time scale of the
  disappearance of the bombarding planetesimals.
\item The mass that we detect through its infrared emission is only
  a minute fraction (about $10^{-5}$) of the mass present. We see the
  gravel but not the very big stones that produce it.
\item Some very young stars lack a disk; some very old stars still
  have a disk: the existence of both groups needs to be explained.
\item Stars in multiple systems retain their remnant disks as often as
  isolated stars.
\end{itemize}

\begin{acknowledgements}

\begin{sloppypar}
  The ISOPHOT data presented in this paper was reduced using PIA,
  which is a joint development by the ESA Astrophysics Division and
  the ISOPHOT consortium. In particular, we would like to thank Carlos
  Gabriel for his help with PIA. We also thank J. Dommanget for
  helpful information on the multiplicity of our stars and the
  referee, R. Liseau for his careful comments. This research has made
  use of the Simbad database, operated at CDS, Strasbourg, France, and
  of NASA's Astrophysics Data System Abstract Service. CD was
  supported by the Stichting Astronomisch Onderzoek in Nederland,
  Astron project 781-76-015.
\end{sloppypar}
     
\end{acknowledgements}

%
%  These Macros are taken from the AAS TeX macro package version 4.0.
%  Include this file in your LaTeX source only if you are not using
%  the AAS TeX macro package and need to resolve the macro definitions
%  in the BibTeX entries returned by the ADS abstract service.
%
%  For more information on the AASTeX macro package, please see the URL
%       http://www.ferberts.com/AAS/aastex.html
%  For more information about ADS abstract server, please see the URL
%       http://adswww.harvard.edu/ads_abstracts.html
%

% Abbreviations for journals.  The object here is to provide authors
% with convenient shorthands for the most "popular" (often-cited)
% journals; the author can use these markup tags without being concerned
% about the exact form of the journal abbreviation, or its formatting.
% It is up to the keeper of the macros to make sure the macros expand
% to the proper text.  If macro package writers agree to all use the
% same TeX command name, authors only have to remember one thing, and
% the style file will take care of editorial preferences.  This also
% applies when a single journal decides to revamp its abbreviating
% scheme, as happened with the ApJ (Abt 1991).

\def\aj{\rm {AJ}}                   % Astronomical Journal
\def\araa{\rm {ARA\&A}}             % Annual Review of Astron and Astrophys
\def\apj{\rm {ApJ}}                 % Astrophysical Journal
\def\apjl{\rm {ApJ}}                % Astrophysical Journal, Letters
\def\apjs{\rm {ApJS}}               % Astrophysical Journal, Supplement
\def\ao{\rm {Appl.~Opt.}}           % Applied Optics
\def\apss{\rm {Ap\&SS}}             % Astrophysics and Space Science
\def\aap{\rm {A\&A}}                % Astronomy and Astrophysics
\def\aapr{\rm {A\&A~Rev.}}          % Astronomy and Astrophysics Reviews
\def\aaps{\rm {A\&AS}}              % Astronomy and Astrophysics, Supplement
\def\azh{\rm {AZh}}                 % Astronomicheskii Zhurnal
\def\baas{\rm {BAAS}}               % Bulletin of the AAS
\def\jrasc{\rm {JRASC}}             % Journal of the RAS of Canada
\def\memras{\rm {MmRAS}}            % Memoirs of the RAS
\def\mnras{\rm {MNRAS}}             % Monthly Notices of the RAS
\def\pra{\rm {Phys.~Rev.~A}}        % Physical Review A: General Physics
\def\prb{\rm {Phys.~Rev.~B}}        % Physical Review B: Solid State
\def\prc{\rm {Phys.~Rev.~C}}        % Physical Review C
\def\prd{\rm {Phys.~Rev.~D}}        % Physical Review D
\def\pre{\rm {Phys.~Rev.~E}}        % Physical Review E
\def\prl{\rm {Phys.~Rev.~Lett.}}    % Physical Review Letters
\def\pasp{\rm {PASP}}               % Publications of the ASP
\def\pasj{\rm {PASJ}}               % Publications of the ASJ
\def\qjras{\rm {QJRAS}}             % Quarterly Journal of the RAS
\def\skytel{\rm {S\&T}}             % Sky and Telescope
\def\solphys{\rm {Sol.~Phys.}}      % Solar Physics
\def\sovast{\rm {Soviet~Ast.}}      % Soviet Astronomy
\def\ssr{\rm {Space~Sci.~Rev.}}     % Space Science Reviews
\def\zap{\rm {ZAp}}                 % Zeitschrift fuer Astrophysik
\def\nat{\rm {Nature}}              % Nature
\def\iaucirc{\rm {IAU~Circ.}}       % IAU Cirulars
\def\aplett{\rm {Astrophys.~Lett.}} % Astrophysics Letters
\def\apspr{\rm {Astrophys.~Space~Phys.~Res.}}
                % Astrophysics Space Physics Research
\def\bain{\rm {Bull.~Astron.~Inst.~Netherlands}} 
                % Bulletin Astronomical Institute of the Netherlands
\def\fcp{\rm {Fund.~Cosmic~Phys.}}  % Fundamental Cosmic Physics
\def\gca{\rm {Geochim.~Cosmochim.~Acta}}   % Geochimica Cosmochimica Acta
\def\grl{\rm {Geophys.~Res.~Lett.}} % Geophysics Research Letters
\def\jcp{\rm {J.~Chem.~Phys.}}      % Journal of Chemical Physics
\def\jgr{\rm {J.~Geophys.~Res.}}    % Journal of Geophysics Research
\def\jqsrt{\rm {J.~Quant.~Spec.~Radiat.~Transf.}}
                % Journal of Quantitiative Spectroscopy and Radiative Trasfer
\def\memsai{\rm {Mem.~Soc.~Astron.~Italiana}}
                % Mem. Societa Astronomica Italiana
\def\nphysa{\rm {Nucl.~Phys.~A}}   % Nuclear Physics A
\def\physrep{\rm {Phys.~Rep.}}   % Physics Reports
\def\physscr{\rm {Phys.~Scr}}   % Physica Scripta
\def\planss{\rm {Planet.~Space~Sci.}}   % Planetary Space Science
\def\procspie{\rm {Proc.~SPIE}}   % Proceedings of the SPIE

\let\astap=\aap
\let\apjlett=\apjl
\let\apjsupp=\apjs
\let\applopt=\ao

\appendix

\section{Observing strategy for minimaps}

\begin{figure}
\centerline{\psfig{figure=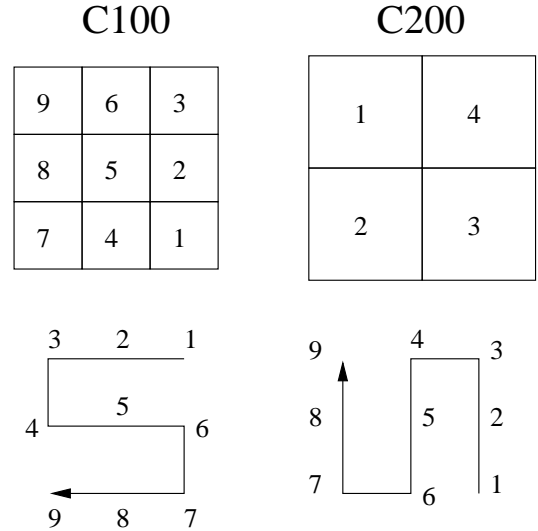,width=7cm,clip=}}
\caption{\label{fig:steps} The upper half labels the different pixels
  of the C100 and C200 detectors seen in projection on the sky. The
  lower half describes the stepping directions: see text.}
\end{figure}

The observations at 60\,\um\  and 170\,\um\  have been taken as minimaps
with the C100 and C200 detector arrays using 3x3 rastersteps ; see
Fig. \ref{fig:steps}. In this figure the upper half shows the
labeling, ``$p$'', of the 9, respectively 4 pixels (detectors) for the
C100 and C200 arrays. The lower diagram gives the numbering, ``$r$'',
of the successive array positions as it moves over the sky; the raster
step is 46 arcseconds for both arrays. Consider first a measurement
with the C100 array.  At raster step $r=1$ the source illuminates
pixel $p=7$; at the next step, $r=2$, the source illuminates $p=4$, at
$r=3$ the source is on $p=1$, etc.. For the C200 measurements $r=1$
has the source on $p=1$; at $r=2$ the source is half on $p=1$ and half
on $p=2$; at $r=3$ the source is on $p=2$, et cetera.

\section{Data reduction}

We used the following procedure to extract the flux. The result of a
minimap measurement is a flux per pixel for each pixel and each raster
position.  Let $f(p,r)$ be the measured flux in pixel $p$ at raster
position $r$.  There are $n_p$ pixels and $n_r$ raster positions.  We
first calculate a flat field correction $f_{\mathrm{flat}}(p)$ for
each pixel by assuming that at one raster position the flux averaged
over all pixels is the same

\newcommand{\oot}{-$\frac{1}{3}$}
\newcommand{\oof}{-$\frac{1}{4}$}
\begin{table}
\caption{\label{tab:raster} Weight factors for minimap flux determination}
\footnotesize
\begin{tabular}{|c|rrrrrrrrr|}
\hline
     & \multicolumn{9}{c|}{Raster position}\\
Pix  &  1   &   2   &   3   &   4   &   5   &   6   &   7   &   8   &   9   \\
\hline
C100 &      &      &      &      &      &      &      &      &      \\
 1 &  0   &   0  &   1  &   0  &   0  & (\oot) & \oot & \oot &   0  \\
 2 & \oot &   0  &   0  &   1  &   0  & (\oot) & \oot &  0   &   0  \\
 3 & \oot & \oot &   0  &   0  &   0  & (\oot) &   0  &  0   &   1  \\
 4 &   0  &   1  &   0  &   0  &   0  & (0)   & \oot & \oot & \oot \\ 
 5 & \oof &   0  & \oof &   0  &   1  & (0)   & \oof &  0   & \oof \\
 6 & \oot & \oot & \oot &   0  &  0   & (0)   &   0  &   1  &   0  \\
 7 &   1  &   0  &   0  & \oot &  0   & (0)   &   0  & \oot & \oot \\
 8 &   0  &   0  & \oot & \oot &  0   & (1)   &   0  &   0  & \oot \\
 9 &   0  & \oot & \oot & \oot &  0   & (0)   &   1  &   0  &  0   \\
\hline
C200 &      &      &      &      &      &      &      &      &      \\
 1 &   1  &   0  &   0  &  0   &  0   &  0   &   0  &  0   & $-1$ \\
 2 &   0  &   0  &   1  &  0   &  0   &  0   & $-1$ &  0   &  0   \\
 3 & $-1$ &   0  &   0  &  0   &  0   &  0   &   0  &  0   &   1  \\
 4 &   0  &   0  &  $-1$&  0   &  0   &  0   &   1  &  0   &  0   \\
\hline
\end{tabular}
\end{table}

\begin{equation}
\label{eq:rooster1}
f_{\mathrm{flat}}(p) =
\frac{\frac{1}{n_p}\sum\limits_{p'=1}^{n_p}\sum\limits_{r'=1}^{n_r}
     f(p',r')}{\sum\limits_{r'=1}^{n_r}f(p,r')}
\end{equation}

Since the individual pixels in the C100/C200 cameras have different
properties, we use the measurements of each pixel to derive a separate
measurement of the source flux.  In order to compute the
background-subtracted source flux, we assign to each raster position
a weight factor $g(p,r)$.  Since the exact point spread function is
not known well enough, we use as on-source measurement the raster
position where the pixel was centered on the source (weight factor 1).
The background measurement is derived by averaging over the raster
postions where the same pixel $p$ was far away from the source (weight
$-1/3$ or $-1/4$).  Raster positions in which the pixel was partially
on the source are ignored (weight 0).  The resulting weight factors
are given in Table~\ref{tab:raster}.  The source flux measured by
pixel $p$ is given by

\begin{equation}
\label{eq:rooster2}
F(p) = \frac{f_{\mathrm{flat}}}{f_{\mathrm{psf}}} 
\sum\limits_{r'=1}^{n_r} g(p,r') f(p,r')
\end{equation}
$f_{\mathrm{psf}}$ is the point spread function correction factor as
given by \citet{laur:00}.  We then derive the flux $F$ and the error
$\sigma$ by treating the different $F(p)$ as independent measurements.

\begin{eqnarray}
\label{eq:rooster3}
F & = & \frac{1}{n_p} F(p)
\end{eqnarray}
\begin{eqnarray}
\label{eq:rooster4}
\sigma & = & \frac{1}{\sqrt{n_p-1}} \sqrt{\sum\limits_{p'=1}^{n_p}
(F(p')-F)^{2}}
\end{eqnarray}

The point spread function is broader than a pixel. We have corrected
for this, using the parameter $f_{\mathrm{psf}}$ given above.  The
correction factor may be too low: \citet{dent:00} show that the disk
around Fomalhaut ($\alpha$ PsA) is extended compared to our
point-spread function. This means that all our 60\,\um\  and 170\,\um\ 
detections are probably somewhat underestimated.

We have ignored pixel 6 of the C100 camera entirely, because its
characteristics differ significantly from those of the other pixels:
it has a much higher dark signal and anomalous transient behaviour.

In the future the characteristics of each pixel will be determined
with increasing accuracy. It may prove worthwhile to redetermine the
fluxes again.

\section{Optical depth of the disk, detection limit and illumination bias}

We discuss how the contrast factor, $C_{60}$, depends on the spectral
type of the star, \Teff, and on the optical depth, $\tau$, of the
disk. The dust grains in the remnant disk are relatively large, at
least in cases where a determination of the grain size has been
possible \citep{bliek:94, arty:89} and the absorption efficiency for
stellar radiation will be high for stars of all spectral types.  The
efficiency for emission is low: the dust particles emit beyond
30\,\um\ and these wavelengths are larger than that of the particles.
We assume that the dust grains are all of a single size, $a$, and
located at a single distance, $r$, from the star. We will introduce
various constants that we will call
$A_i$, $i=0-6$.
\begin{equation}
\label{eq:c1}
C_{60}\equiv
\frac{L_{\nu,\mathrm{d}}}{L_{\nu,*}}=
\frac{L_{\nu,\mathrm{d}}}{L_{d}}\cdot
\frac{L_\mathrm{d}}{L_{*}}\cdot \frac{L_{*}}{L_{\nu,*}}.
\end{equation}

First we determine $L_{\nu,{\mathrm{d}}}$, the luminosity of the disk
at the frequency $\nu$, and $L_{\mathrm{d}}$, the total luminosity of
the disk: 
\begin{equation}
\label{eq:c2}
L_{\nu,{\mathrm{d}}}=N 4\pi^2 a^2 Q_\nu B_\nu(T_{\mathrm{d}})
\end{equation}
and
\begin{equation}
\label{eq:c3}
L_{\mathrm{d}}= N 4\pi^2 a^2\int_0^\infty Q_\nu B_\nu(T_{\mathrm{d}}) d\nu.
\end{equation}
We consider dust emission at 60\,\um; $B_\nu(T)$ can be approximated by the 
Wien-equation. We thus write:
\begin{equation}
\label{eq:c4}
L_{\nu,\mathrm{d}}= A_0 \exp(-\frac{240\,K}{T_{\mathrm{d}}})
\end{equation}

Define the average absorption efficiency:
\begin{equation}
\label{eq:c5}
Q_{\mathrm{ave}}(T_{\mathrm{d}})\equiv\pi{\int_{0}^{\infty}}{Q_\nu B_\nu}
(T_{\mathrm{d}}) d\nu /(\sigma T_{\mathrm{d}}^4).
\end{equation}
For low dust-temperatures $Q_{\mathrm{ave}}$
can be approximated by $Q_{\mathrm{ave}}=A_1 \Tdust^{\alpha}$ with
$\alpha\approx 2$ \citep{natt:76}
and thus
\begin{equation}
\label{eq:c6}
L_{\mathrm{d}}= A_2 T_{\mathrm{d}}^6.
\end{equation}\\

Second, we determine the stellar luminosity,  $L_{\nu,*}$, at
frequency $\nu$ and the total stellar luminosity, $L_*$, both
by ignoring the effects of dust, that is the luminosity at the
photospheric level. The photospheric emission is approximated
by the Rayleigh-Jeans equation:
\begin{equation}
\label{eq:c7}
L_{\nu,*}=\frac{\pi B_\nu(T_{\mathrm{eff}})}
{\sigma T_{\mathrm{eff}}^4} L_*= \frac{A_3 L_*}{T_{\mathrm{eff}}^{3}}
\end{equation}
The stellar luminosity for main sequence stars of spectral type A0--K5
can be approximated within 30\% by: 
\begin{equation}
\label{eq:c8}
L_*=A_4 T_{\mathrm{eff}}^{8.2}
\end{equation}\\

Third, we determine the relation between $T_{\mathrm{d}}$ and
$T_{\mathrm{eff}}$. For photospheric temperatures $Q_{\mathrm{ave}}$
is independent of \Teff. The energy absorbed by a grain is thus
$\propto L_*\propto T_{\mathrm{eff}}^{8.2}$. The energy emitted is
$\propto T_{\mathrm{d}}^6$. Because the energy emitted equals the
energy absorbed we conclude that $T_{\mathrm{eff}}= A_5
T_{\mathrm{d}}^{6/8.2}$.\\

Combining these results we find
\begin{equation}
\label{eq:c9}
C_{60}=
{A_6}\cdot\frac{1}{T_{\mathrm{d}}^{3.8}}\cdot
\exp(-\frac{240}{T_{\mathrm{d}}}).
\end{equation} 

\begin{figure}
\psfig{figure=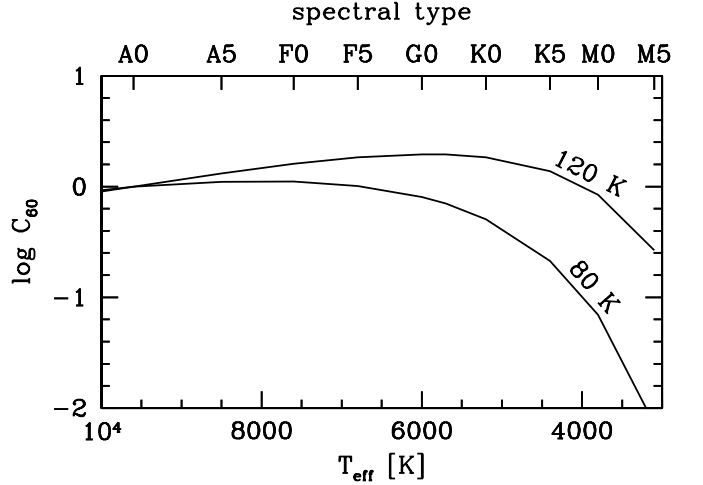,width=8.8cm,clip=}
\caption{\label{fig:contrast-analytically} The contrast factor $C_{60}
  \equiv F_\nu^{\mathrm{d}}/F_\nu^{\mathrm{pred}}$ for a ring of dust
  at a distance of 50 AU from the star and for stars of different
  effective temperature. We assumed $T_{\mathrm{d}}$ to be 80 or 120K
  for an A0 star.}
\end{figure}

We have calculated $C_{60}$ without making the various approximations
in Eqs. \ref{eq:c1} through \ref{eq:c7}: Fig.
\ref{fig:contrast-analytically} shows the results. The results are
valid if the distance dustring/star is the same (50\, AU) for all stars
irrespective of the spectral type.

In the figure we assume that for an A0-star $C_{60}$ has the value 1
and $T_{\mathrm{d}}= 80$ or $120\,$K. For a star of later spectral
type, the dust will be cooler and will emit less energy (see
Eq.\,\ref{eq:c6}), but since $\lambda= 60\mu$m is at the Wien-side of
the black body curve, the emission at 60\um\ will increase- see
Eq.\,\ref{eq:c4}. The consequence is that $C_{60}$ remains constant
for A-, F- and early G-type stars.  For late G- and for K- and M-type
stars the dust becomes too cold to be detected at 60\um.  Only
photometry at longer wavelengths will ultimately be able to detect
such very cold disks.

\section{Determination of $\tau$ from observed fluxes}

For most of our stars we have only a detection of the disk at 60\um.
To calculate the optical depth and the mass of the disk we need an
estimate of the disk emission integrated over all wavelengths.
If the dust around an A star has a temperature of \Tdustref, the
stars of later types will have lower dust temperatures.  Numerical
evaluation shows that, assuming constant distance between the star and
the dust, the following relation is a good approximation

\begin{equation}
\label{eq:16}
\Tdust = \Tdustref - 
\frac{\Tdustref}{\Tstarref-2000}
(\Tstarref-\Tstar) \quad .
\end{equation}

With this estimate of the dust temperature, a single flux
determination is sufficient to determine the the fractional
luminosity.  We have used this method to determine estimates of $\tau$
independently from all wavelength where we have determined an excess.
We used \Tstarref = 9600 K and \Tdustref = 80 K, values that agree
with those measured for Vega.

\end{document}